\documentclass[english,aps,jmp,showpacs,titlepage,authoryear,url,12pt]{revtex4}

\usepackage{amssymb}
\usepackage{amsmath}
\usepackage{mathtools}
\usepackage{epigraph}
\usepackage{eucal}
\usepackage[dvips]{graphics}
\usepackage{epsfig}
\usepackage{bm}
\usepackage{emerald}
\usepackage{babel}
\usepackage{slantsc}
\usepackage{array}
\usepackage{hyperref}
\usepackage[sort&compress]{natbib}
\usepackage{url}

\def\st{\scriptstyle}

\def\be{\begin{equation}}
\def\ee{\end{equation}}
\def\ba{\begin{eqnarray}}
\def\ea{\end{eqnarray}}
\def\bsu{\begin{subequations}}
\def\esu{\end{subequations}}

\def\H{{\mathcal H}}

\def\M{{\mathcal M}}

\def\a{\alpha}
\def\b{\beta}
\def\g{\gamma}     \def\G{\Gamma}
\def\d{\delta}

\def\l{\lambda}
\def\m{\mu}
\def\n{\nu}
\def\o{\omega}   \def\O{\Omega}

\def\s{\sigma}		
\def\t{\tau}

\def\lab{\label}
\def\pd{\partial}
\def\le{\left}
\def\ri{\right}

\def\mm{{\mathtt f}}

\def\phys{$(\M,g^{*}_{\a\b},\G^{*\a}_{\b\g})\;$}
\def\conf{$(\M,\mm_{\a\b},{\cal G}^\a_{\b\g})\;$}
\def\tang{$(T_\M,\mm_{\a\b},{\cal G}^\a_{\b\g})\;$}

\begin{document}
\title{Einstein's equivalence principle in cosmology.}
\author{Sergei M. Kopeikin}
\affiliation{Department of Physics \& Astronomy, University of Missouri, 322 Physics Bldg., Columbia, MO 65211, USA}
\email{kopeikins@missouri.edu}
\date{\today}

\begin{abstract}
We study physical consequences of the Einstein equivalence principle (EEP) for a Hubble observer in the conformally-flat Friedmann-Lema\^itre-Robertson-Walker (FLRW) spacetime covered by global coordinates $y^\a=(\eta,x^i)$ where $\eta$ is a conformal time. In accordance with EEP, we introduce the local inertial coordinates $x^\a=(x^0,x^i)$ in the vicinity of the observer's world line with the help of a coordinate diffeomorphism, $y^\a=y^\a(x^\b)$, called the special conformal transformation that respects the local conformal equivalence between the tangent spacetime of the observer and the FLRW background manifold. The local inertial metric is Minkowski flat, $\mm_{\a\b}={\rm diag}(-1,1,1,1)$ and materialized by a congruence of time-like geodesics of static observers being at rest with respect to the local spatial coordinates $x^i$. The static observers are equipped with the ideal clocks measuring their own proper time which is synchronized with the proper time $\t$ measured by the Hubble observer. The local inertial metric is used for physical measurements of spacetime intervals rendered with the ideal clocks and the ideal rigid rulers. The special conformal transformation preserves null geodesics but does not keep invariant time-like geodesics. Moreover, it makes the rate of the local time coordinate $x^0=x^0(\eta)$, as a function of the conformal time $\eta$ taken on the world line of a particle, be dependent on the velocity of the particle. Hence, it is not possible to rich the uniform parametrization of the world line of a static observer and that of the light geodesics with one and the same parameter of the proper time taken on the world line of the static observer. If we denote the proper time of the static observer $\tau$ and the parameter along the light geodesic $\lambda$, they have to differ by the conformal factor $a(\t)$ of FLRW metric as a consequence of the special conformal diffeomorphism. The most convenient way, then, to study the local propagation of light in cosmology is achieved in terms of the {\it optical} metric $g_{\a\b}$ induced by projecting FLRW metric on the light cone which characteristics are parametrized by the proper time $\t$ of the Hubble observer. The local optical metric is orthogonal but not flat for it depends on the scale factor $a(\t)$ of FLRW universe that can be interpreted as an appearance of a weak violation of EEP for photons if we pretend to treat the background spacetime as asymptotically-flat. The importance of this observation is that, at least, some of the local astronomical experiments conducted with freely-propagating electromagnetic waves are to be sensitive to the Hubble expansion, if our mathematical model conforms to reality. We analyse some of these experiments and demonstrate that the Hubble constant, $H=\dot a/a$, can be measured within the solar system by means of high-precision spacecraft Doppler tracking as a systematic {\it blue} shift of frequency of radio waves circulating in the Earth-spacecraft radio link. We also analyze the behavior of the standing wave in a microwave resonator cavity on cosmological manifold and show that contrary to the freely-propagating radio wave, the standing wave in the cavity is insensitive to the Hubble expansion. 
\end{abstract}
\maketitle

\section{Introduction}

Modern physics is intensively looking for the unified field theory that might explain the origin of the universe and the underlying fundamental nature of spacetime and elementary particles  \citep{baryshev_2012ASSL}. This work requires deeper understanding of the theoretical and experimental principles of general relativity. It is challenging to find a new type of experiments that broaden the current knowledge. An appealing problem is to examine a presumable link between the local gravitational phenomena and the global cosmological expansion of the universe that is to test the foundational basis of the Einstein equivalence principle (EEP) in application to a conformal metric with a time-dependent scale factor. 

It is commonly accepted that EEP in general relativity is universally valid because gravitational field in general relativity has a pure geometric nature \citep{2006LRR.....9....3W}. Indeed, it is always mathematically possible to find a local diffeomorphism which reduces manifold's metric to a Minkowski metric in a sufficiently small neighborhood of a time-like world line of observer if tidal forces are neglected. This mathematical fact was a clue that led Einstein to formulation of his general principle of relativity (also known as the principle of covariance \citep{kopeikin_2011book}) and, later on, to the discovery of general relativity as a physical theory of gravitational field \citep{einstein1961relativity,Einstein_P1916}. 

The apparent mathematical nature of EEP caused some physicists to deny its physical significance \citep{Norton_1993}. The present paper neither shares this extremal point of view nor confronts the solid mathematical foundation of EEP. We focus on a more subtle aspects of EEP, namely, 
\begin{enumerate}
\item comparison of the inertial motion of test particles along time-like and light geodesics on conformal manifold of cosmological background metric considered from the local point of view of a Hubble observer,
\item derivation of experimental consequences that can be used for testing the Hubble law in the local experiments thus, confirming the universality of cosmological metric at different scales from clusters of galaxies down to the solar system and terrestrial laboratory.  
\end{enumerate} 

So far, all gravitational experiments in the solar system have been interpreted under a rather natural assumption that the background spacetime geometry is asymptotically-flat \citep{will_1993}. On the other hand, theoretical and observational cosmology postulates that the background spacetime is described by the Friedmann-Lema\^itre-Robertson-Walker (FLRW) metric 
\be\lab{1}
ds^2=-dt^2+{R}^2(t)\le(1+\frac14 kr^2\ri)^{-2}\d_{ij}dy^idy^j\;,
\ee
where $t$ is the universal cosmological time, $y^i$ are the global isotropic coordinates, Latin indices $i,j,k,..$ take values $1,2,3$,  $k=\{-1,0,+1\}$ defines a curvature of space, and the scale factor $R(t)$ is a function of time found by solving Einstein's equations \citep{2005pfc..book.....M,weinberg_2008}. Here and everywhere else, we use a geometric system of units in which the fundamental speed, $c$, and the gravitational constant, $G$, are set equal to unity: $G=c=1$. Usually, FLRW metric is considered as a result of averaging over rather large spatial volumes and, hence, is applicable only globally while a local geometry is approximated by the Minkowski spacetime. If this is true, there must be a smooth matching between the global and local geometries allowing us to eliminate any cosmological effect associated with the Hubble expansion in a local frame of a Hubble observer. This is exactly the realm of EEP which we shall explore in the present paper.

In what follows, we admit $k=0$ in accordance with observations \citep{wmap_2011} but it does not limit the results of this paper since the effects associated with the space curvature, $k=\pm 1$, are beyond the linearised Hubble approximation which is applied in the present paper (that is we consider only terms being linear with respect to the Hubble constant). FLRW metric (\ref{1}) is not asymptotically-flat and has a non-vanishing spacetime curvature tensor $\bar R_{\a\b\g\d}$ where the Greek indices $\a,\b,\g,...$ take values $0,1,2,3$. Nonetheless, the Weyl tensor of FLRW metric, $\bar C_{\a\b\g\d}\equiv 0$. Hence, (\ref{1}) can be reduced to a conformally-flat metric for any value of space curvature $k$ \citep{2007JMP....48l2501I}. When $k=0$, it is achieved by transforming the cosmological time $t$ to a conformal time, $\eta=\eta(t)$, defined by an ordinary differential equation
\be\lab{oka1}
dt=a(\eta)d\eta\;,
\ee
where the scale factor, $a(\eta)\equiv R[t(\eta)]$. The time transformation (\ref{oka1}) brings the cosmological metric (\ref{1}) into the conformally-Minkowskian form
\be\lab{q1}
ds^2=a^2(\eta)\mm_{\a\b}dy^\a dy^\b\;,
\ee
where $y^\a=(y^0,y^i)=\le(\eta,y^i\ri)$ are the global conformal coordinates, $\mm_{\a\b}={\rm diag}(-1,1,1,1)$ is the Minkowski metric. The reason for denoting the Minkowski metric as $\mm_{\a\b}$ instead of the conventional $\eta_{\a\b}$ is to prevent a confusion with the standard notation for the conformal time $\eta$. 

Equation (\ref{q1}) reveals that FLRW spacetime has two metrics -- the physical metric  $\bar g_{\a\b}\equiv a^2(\eta)\mm_{\a\b}$ and an auxiliary flat metric $\mm_{\a\b}$ -- both residing on the global cosmological manifold $\M$. Metric $\bar g_{\a\b}$ defines the metric relationships in physical spacetime $\le(\M,\bar g_{\a\b}\ri)$ while the Minkowski metric $\mm_{\a\b}$ determines the metric properties of the conformally-flat Minkowski spacetime $(\M,\mm_{\a\b})$. It is important to emphasize that the conformally-flat spacetime {\it is not physical} and serves merely as a mathematical device to specify the global geometric properties of FLRW universe \citep{waldorf}.

The two metrics generate two affine connections on FLRW manifold \citep{waldorf}. The first affine connection is introduced as the Christoffel symbol, $\bar\G^{\a}_{\b\g}$, made out of the metric $\bar g_{\a\b}$. The second (flat) connection is defined as the Christoffel symbol, ${\cal G}^\a_{\b\g}$, being compatible with the Minkowski metric $ \mm_{\a\b}$. We denote the physical spacetime $(\M,\bar g_{\a\b},\bar\G^{\a}_{\b\g})$ and the conformally-flat spacetime $(\M,\mm_{\a\b},{\cal G}^\a_{\b\g})$.
The two connections are interrelated 
\be\lab{qq2}
\bar\G^{\a}_{\b\g}= {\cal G}^\a_{\b\g}+C^\a_{\b\g}\;,
\ee
where  
\be\lab{qq3}
C^\a_{\b\g}=\d^\a_\b\pd_\g\ln a+\d^\a_\g\pd_\b\ln a-\eta_{\b\g}\eta^{\a\m}\pd_\m\ln a\;,
\ee
is a tensor, called the Weyl connection \citep{schmidt_1987}, and $\pd_\a\equiv\pd/\pd y^\a$. Partial derivative $\pd_\a \ln a=-H\bar u_\a$ where $H=d\ln a/dt$ is the Hubble constant, $ \bar u_\a= \bar g_{\a\b} \bar u^\b$, and $\bar u^\a$ is a tangent vector to the time-like geodesic worldline of a Hubble observer normalized such that $\bar g_{\a\b} \bar u^\a \bar u^\b=-1$. In the global coordinates $(t,y^i)$ of FLRW metric (\ref{1}) the four-velocity of the Hubble observer $\bar u^\a=(a^{-1},0,0,0)$. 

The rule of the parallel transport on the physical FLRW manifold \phys is defined by picking up a specific value of the flat connection in \conf. It is postulated that the Christoffel symbols ${\cal G}^\a_{\b\g}(y)=0$ in the global conformal coordinates, $y^\a$, and in any other coordinates that are related to $y^\a$ by the Poincar\'e transformation \cite{waldorf}. This choice of the flat connection yields, $\bar\G^{\a}_{\b\g}(y)=C^\a_{\b\g}$, so that the equations of geodesics in \phys become fully determined. The reader should remember that in arbitrary (curvilinear) coordinates, $x^\a$, introduced on the cosmological manifold, $\M$, the flat connection ${\cal G}^\a_{\b\g}(x)$ is not nil and the general equation (\ref{qq2}) must be applied to find our $\bar\G^{\a}_{\b\g}(x)$ in the curvilinear coordinates. In any case, the curvature tensor of FLRW manifold is defined by the connection $ C^{\a}_{\b\g}(x)$ only \citep{waldorf}
\be\lab{a0}
\bar R^{\a}{}_{\b\m\n}=C^{\a}{}_{\b\n,\m}-C^{\a}{}_{\b\m,\n}+C^{\a}{}_{\m\g}C^{\g}{}_{\b\n}-C^{\a}{}_{\n\g}C^{\g}{}_{\b\m}\;.
\ee
and can be calculated directly in terms of the scale factor of FLRW metric.

According to Einstein's general relativity and the definition of FLRW metric,
the cosmological time $t$ is a physical proper time $\t$ of the Hubble observer and can be measured with the help of the observer's atomic clock while the conformal time $\eta$ is a convenient coordinate parameter which is calculated from the clock's reading but cannot be measured directly \citep{2005pfc..book.....M,weinberg_2008}. Typically, the cosmological metric (\ref{q1}) is applied to describe the properties of spacetime on the scale of galaxy clusters and larger. On small scales of the size of the Milky Way, the solar system and terrestrial lab, the background spacetime is believed to be flat with any cosmological effect being strongly suppressed. Nevertheless, the question remains open: if we admit FLRW metric to be valid on any scale, can the cosmological expansion be detected in local gravitational experiments? Review article \citep{2010RvMP...82..169C} gives a negative answer to this question. Our recent paper \citep{Kopeikin_2012eph} on celestial dynamics in an expanding universe challenges this opinion. The present paper supports our previous analysis by analysing EEP from the point of view of measurements made locally by a Hubble observer.    

Similarly to \citep{2010RvMP...82..169C,Kopeikin_2012eph}, we postulate that FLRW metric (\ref{q1}) is a physical metric not only in cosmology but for the  description of the local physics as well. It describes the background spacetime geometry in the global coordinates $y^\a$ on all scales spreading up from the cosmological horizon to the solar system and down to a local observer. The small parameter in the approximation scheme used in the present paper, is the product of the Hubble constant, $H$, with the interval of time used for physical measurements. All non-linear terms of the quadratic order with respect to the small parameter (formally, the terms being quadratic with respect to $H$) will be systematically neglected because of their smallness. 

The collection of the linear-order terms in the expansion of cosmological equations with respect to the Hubble constant, is termed the linearised Hubble approximation. In this approximation we do not consider the effects of cosmological curvature - only the metric tensor and the affine connection (the first derivatives of the metric tensor along with the Weyl connection) are involved into the mathematical treatment. This is exactly the realm of EEP.
We shall also neglect all special-relativistic and post-Newtonian effects which must be included to the realistic data analysis of observations in the solar system. These effects are described in \citep{kopeikin_2011book,2013strs.book.....S}, and can be easily accounted for, if necessary.

We introduce the reader to the concepts associated with the Einstein equivalence principle in section \ref{eepr} and discuss construction of the local inertial frame in section \ref {preq}. The inertial and optical metrics on the tangent spacetime in local coordinates are derived in sections \ref{sec4} and \ref{opm123} respectively. Section \ref{opm123} also explains why we have to use the optical metric locally for physical interpretation of light propagation. The optical metric is used in section \ref{lglc} to derive equation of light geodesics in local coordinates. We solve these equations in section \ref{rr1} and employ them for investigation of observability of cosmological effects in the solar system. We also explore the case of a resonant cavity in section \ref{kk4d} to see if the local cosmological expansion can be detected by precise time metrology. Final comments and discussion are given in section \ref{opwd}.

\section{Einstein's principle of equivalence}\lab{eepr}

A thorough  treatment of the local astronomical measurements on cosmological manifold inquires a scrutiny re-examination of Einstein's equivalence principle (EEP) which states: ``In a given gravitational field, the outcome of {\it any} local, non-gravitational experiment is independent of the freely-falling experimental apparatus' velocity, of where and when in the gravitational field the experiment is performed and of experimental technique applied'' \citep{Rohrlich1963169}. Mathematical interpretation of EEP suggests universality of local geometry in the sense that at each point on a spacetime manifold with an arbitrary gravitational field, it is possible to chose the local inertial coordinates such that, within a sufficiently small region of the point in question, {\it all} laws of nature take the same form as in non-accelerated Cartesian coordinates \citep{Harvey1964383,kopeikin_2011book}. 

It is generally accepted that EEP is applicable to any kind of spacetime manifold, in particular, to the manifold of FLRW universe \citep{2010RvMP...82..169C} which is described by a conformally-flat metric (\ref{q1}). We noticed \citep{Kopeikin_2012eph,kopeikin_2013} that due to the conformal nature of FLRW manifold, the propagation of light in local coordinates differs from its propagation in the Cartesian coordinates of flat spacetime. This paper continues to study this cosmologically-related phenomenon with the aim to reach a more comprehensive understanding of the influence of the Hubble expansion on the local motion of photons. It draws to the conclusion that EEP is violated for photons propagating on cosmological manifold due to the local conformal equivalence of FLRW and Minskowsky metrics. The conformal equivalence of the metrics preserves the equivalence between light geodesics but it does not suggest the conformal equivalence between time-like geodesics \citep{waldorf}. Transformation to the local frame on cosmological manifold cannot make the affine connection nil simultaneously along time-like and light geodesics on FLRW manifold. We shall solve this "mistery" in the following sections of the present paper. 

According to EEP a Hubble observer carries out the local inertial coordinates (LIC), $x^\a=(x^0,x^i)$ such that the physical metric $\bar g_{\a\b}\equiv a^2(\eta)\mm_{\a\b}$, given by (\ref{q1}), is diffeomorphic to the Minkowski metric, $\mm_{\a\b}$, 
with the affine connection being nil on the observer's world line \citep{hongya:1920,hongya:1924,2005ESASP.576..305K,2007CQGra..24.5031M}. EEP also asserts that the worldlines of freely falling (electrically-neutral) test particles and photons are geodesics of the physical metric $\bar g_{\a\b}$ with an affine parametrization. It presumes that in LIC, the geodesic equations of motion of {\it all} test particles -- massive and massless -- can be written down as 
\be\lab{bq1}
\frac{d^2x^\a}{d\sigma^2}=0\;,
\ee
where $\sigma$ is the affine parameter along the geodesic, under condition that the tidal (caused by the Riemann curvature of FLRW spacetime) force is neglected \citep{nizim}. The parameter $\sigma$ can be chosen to be equal to the coordinate time $x^0$ of LIC which, in its own turn and under conventional choice of units, can be made equal to the proper time $\t$ of the Hubble observer located at the origin of the local coordinates. This can be done for any type of geodesics - time-like and light-like. Indeed, if we accept that (\ref{bq1}) is true, then, it tells us that $x^0=\t$ is a linear function of $\sigma$ and, hence, a re-parametrization of (\ref{bq1}) with $\t$ transforms it to
\be\lab{bq1a}
\frac{d^2x^\a}{d\t^2}=0\;.
\ee

This equation, when applied to a photon and solved, yields the photon's world line $x^0=\t,\;x^i=k^i\t$, where $k^i$ is a unit vector in the direction of the photon's propagation and we assumed that light passes through the origin of the LIC at instant $\t=0$ which fixes the integration constants. It establishes a one-to-one relationship between the spatial coordinates $x^i$ of the photon and the proper time $\t$ of the observer at the origin of LIC, which is a directly measurable quantity. For this reason, the light cone solution of (\ref{bq1a}) is one of the basic equations of radar astronomy serving for measuring distances and velocities of spacecraft and planets in the solar system \citep{kopeikin_2011book,2013strs.book.....S} \footnote{Realistic measurement requires accounting for the post-Newtonian relativistic corrections in (\ref{bq1a}) which is beyond the scope of the present paper but can be found, for example, in \citep{kopeikin_2011book,2013strs.book.....S}.}. 

Neither equation (\ref{bq1}) nor (\ref{bq1a}) show the presence of the Hubble constant, $H$, which is incorporated to the Weyl connection $C^\a_{\b\g}$ of FLRW manifold. It led scientists to believe that EEP cancels out all cosmological effects of the linear order of $O(H)$ that prevents astronomers to observe them in the solar system' experiments \citep{2010RvMP...82..169C}. The tidal cosmological forces, which are of the quadratic order of $O(H^2)$, are not cancelled out in equations of motion but they negligibly small and cannot be observed  \citep{2005ESASP.576..305K,2007CQGra..24.5031M,2010RvMP...82..169C}, so we are not discussing them over here. 

The present paper analyses the applicability of EEP to the important case of a conformal manifold of FLRW universe and suggests that equation (\ref{bq1a}) is not hold for photons in LIC where the time coordinate $x^0$ is measured as the proper time $\t$ of the Hubble observer located at the origin of the LIC. It turns out that though the physical metric, $\bar g_{\m\n}(x)=\mm_{\a\b}={\rm diag}\le[-1,1,1,1\ri]$ in the local coordinates $x^\a=(\t,x^i)$, it does not correspond to the uniform propagation of light with constant velocity. This non-uniform propagation of light in the local coordinates may be interpreted as a violation of EEP for photons. 

It turns out that if we use the local (physical) coordinates $x^\a=(\t,x^i)$, light propagates along geodesics of the {\it optical} metric ${g}_{\a\b}={\rm diag}\le[-a^2(\t),1,1,1\ri]$ where $a(\t)=1+H\t$ is the scale factor of FLRW metric, and we restrict ourselves with the linearised Hubble approximation. It means that equation (\ref{bq1a}) must be amended with an additional term corresponding to the non-vanishing components of the affine connection of the optical metric. We derive such a local equation of motion for photons in section \ref{opm123}. It includes a term of the linear order of $O(H)$ which makes the measurement of the Hubble expansion in the solar system feasible in the local experiments like the Doppler tracking of spacecraft.  

We should emphasize that we do not introduce any new metric "by hands". The metric used is one and the same - the FLRW metric, but it is written in different coordinates which take into account the re-scaling of space and time and the correspondence between physical observables and coordinates.  This is a subtle point of the whole discussion which should be followed step by step. All conclusions about physics which we derive in the present paper, follow directly from the rigorous mathematical transformations of the original FLRW metric.

The line of our reasoning is as follows. Physical spatial coordinates in cosmology are $x^i=R(t)y^i$. Substituting this transformation to FLRW metric (\ref{1}) with $k=0$, transforms it to 
\be\lab{tgy2}
ds^2=-\le(1+H^2{\bm x}^2\ri)dt^2-2Hx^idx^idt+\d_{ij}dx^idx^j\;,
\ee
This is still FLRW metric writen down in the coordinates $(t,x^i)$. The time $t$ coincides with the proper time $\t$ of the Hubble observer located at the origin of the local coordinates $x^i=0$. Indeed, equation (\ref{tgy2}) for such an observer is reduced to, $-d\t^2=ds^2=-dt^2$, yielding $\t=t$. We want to describe motion of photons emitted by the Hubble observer at the origin of the local coordinates at time $\t=0$. The world line of photons is given by the condition $ds=0$ and, in the first approximation, is just a straight line $x^i=k^i\t$ where $k^i$ is a unit vector in the Euclidean sense $\d_{ij}k^ik^j=1$. In order to get the second order correction to the motion of photons, we substitute the first order light geodesic to the perturbed (by the Hubble expansion) components of FLRW metric, which yields the FLRW metric on the light-ray hypersurface in the local (physical) coordinates
\be\lab{tgy3}
ds^2 =-(1+H\t)^2d\t^2+\d_{ij}dx^idx^j\;.
\ee
This is not another metric but the same FLRW metric taken on the light cone in the local coordinates used by the Hubble observer for physical measurements. Noticing that in the linearized Hubble approximation the scale factor $R(t)=1+Ht=1+H\t$, we arrive to FLRW metric on the null cone
\be\lab{tgy4}
ds^2=-a^2(\t)d\t^2+\d_{ij}dx^idx^j\;,
\ee
where we have used a conventional notation for the scale factor $a(\t)\equiv R(t)$.
We call the metric (\ref{tgy4}) the {\it optical} metric. Light propagates in the local coordinates $(\t,x^i)$ along geodesics of the optical metric which are determined by $ds=0$. The reader can see that there is no "another" metric over here, only FLRW metric properly transformed to the local coordinates on the light-ray (null) hypersurface is used. The {\it optical} metric is not a new geometric object that appears "in addition" to FLRW metric. It was obtained by projecting FLRW metric on the null hypersurface in the local (physical) coordinates, nothing else. We explain these transformations and their physical consequences for observations in the solar system in more detail in the text which follows.

\section{The local inertial coordinates}\lab{preq}

In order to interpret the local astronomical measurements (like radar ranging, spacecraft Doppler tracking, etc..) we have to build the local inertial coordinates (LIC) in the neighbourhood of a time-like world line of observer. We focus on building LIC in the vicinity of a Hubble observer which is by definition has constant spatial coordinates $y^i={\rm const.}$ of the FLRW metric and moves along a time-like geodesic worldline \citep{waldorf}. Real observers move with respect to the Hubble flow and experience gravitational forces from massive bodies of the solar system. Therefore, construction of LIC for a real observer requires to work out additional coordinate transformations which can be found in \citep{kopeikin_2011book,2013strs.book.....S} but they are not a matter of concern of the present paper.  

Let us put the Hubble observer at the origin of LIC, $x^i=0$, which worldline coincides, then, with the time-like geodesic of the observer. The Hubble observer carries out an ideal clock that measures the parameter of the observer's worldline which is the observer's proper time $\t$. The proper time $\t$ of the Hubble observer coincides with the cosmological coordinate time $t$ in (\ref{1}), that is $\t=t$. EEP suggests that in a small neighbourhood of the worldline of the observer (called a tangent spacetime) there exists a diffeomorphism from the global, $y^\a$, to local, $x^\a$, coordinates such that the physical metric $\bar g_{\a\b}=a^2(\t)\mm_{\a\b}$ is transformed to the Minkowski metric, $\mm_{\a\b}$, as follows
\be\lab{c2}
a^2(\t)\mm_{\m\n}\frac{\pd y^\m}{\pd x^\a} \frac{\pd y^\n}{\pd x^\b}=\mm_{\a\b}\;,
\ee
where all tidal terms of the order of $O(H^2)$ have been omitted as negligibly small. In the tangent spacetime where (\ref{c2}) is valid, the physical spacetime interval (\ref{q1}) written down in LIC, reads
\be\lab{act6}
ds^2=\mm_{\a\b}dx^\a dx^\b\;,
\ee
where $\mm_{\a\b}$ is understood as the physical metric $\bar g_{\a\b}(x)$ expressed in the local coordinates. We emphasize that equation (\ref{c2}) is a local diffeomorphism that should be neither confused with a class of the conformal changes of metrics worked out for applications to the initial value problem in general relativity \cite{Mielke_1977} nor with the Bondi-Metzner-Sachs group of coordinate transformations at null infinity of asymptotically-flat spacetime \cite{BMS_group}.    

Equation (\ref{c2}) looks similar to the special conformal transformation establishing a conformal isometry of the Minkowsky metric  \citep{1966PhRv..150.1183K,cft_intro}
\be\lab{c5}
\O^{2}(x)\mm_{\m\n}\frac{\pd y^\m}{\pd x^\a} \frac{\pd y^\n}{\pd x^\b}=\mm_{\a\b}\;,
\ee
where
\ba\lab{c6}
\O(x)&=&\mm_{\a\b}\le(b^2x^\a-b^\a\ri)\le(b^2x^\b-b^\b\ri)b^{-2}\\\nonumber
&=&1-2b_\a x^\a+b^2 x^2\;,
\ea
is a conformal factor, $b^\a$ is a constant four-vector yet to be specified, $x^2\equiv\mm_{\a\b}x^\a x^\b$, and
$b^2\equiv\mm_{\a\b}b^\a b^\b$. The group of the conformal isometry (\ref{c5}) of flat spacetime depends on 15 parameters including the Poincar\'e
transformations (10 parameters), dilatation (1 parameter), and the special conformal transformation (4 parameters). The special conformal transformation
includes inversions and translations, and is defined by equation \citep{1962AnP...464..388K,cft_intro}
\be\lab{mu7a}
\frac{y^\a}{y^2}=\frac{x^\a}{x^2}-b^\a\;,
\ee
that is equivalent to
\be\lab{c4}
y^\a=\frac{x^\a-b^\a x^2}{\O(x)}\;.
\ee
All operations of rising and lowering indices are completed with the Minkowski metric $\mm_{\a\b}$ and/or $\mm^{\a\b}$.  

Let us assume for simplicity that the origin of LIC, $x^i=0$, coincides with the point having the global spatial coordinates, $y^i=0$. As the background manifold is assumed to be analytic, equation
(\ref{c2}) should match (\ref{c5}) in a small neighbourhood of the origin of the LIC. The matching can be achieved by demanding the scale factor of the FLRW metric, $a(\eta(x))=\Omega(x)$. This equality is 
valid in arbitrary cosmological model if we discard the curvature terms being proportional to $\sim H^2$ and/or $H'$. Indeed, for small values of the conformal time $\eta$ we have,
$a(\eta)=a(0)+a'(0)\eta+a''(0)\eta^2/2+O(\eta^3)$, where we assume that the present epoch corresponds to $\eta=0$ in the conformal time, and the prime
denotes the time derivative, $a'=da/d\eta$, etc. We normalize the scale factor at the present epoch to $a(0)=1$. Then, at the present epoch the Hubble
constant $H=a'(0)$. The second time derivative of the scale factor, $a''=H'+2H^2$, and we drop it off as being negligibly small. Assuming that the constant vector, $b^\a=O(H)$, we approximate the conformal factor, $\O(x)=1-2b_\a x^\a+O(b^2x^2)$. Taking into account that (\ref{c4}) yields $\eta=x^0+O(b)$, and equating $\O(x)$ in (\ref{c6}) to the Taylor expansion of the scale
factor $a(\eta)$, we find out that vector $b^\a=(H/2)u^\a=(H/2,0,0,0)$. It is directed along the four-velocity $u^\a=(1,0,0,0)$ of the Hubble observer and is time-like. 

The reader may notice that the special conformal transformation has a singular point, $x^\a=-b^\a/b^2$, that goes over to $\t=2/H$. It means that the special conformal diffeomorphism (\ref{c4}) is approximately limited in time domain by the Hubble time, $T_H=1/H$, calculated for the present value of the Hubble parameter $H\simeq 2.3\times 10^{-18}$ s$^{-1}$. However, because the LIC have been derived by matching of the local flat metric with the global FLRW metric, the period of time for which the local inertial frame is really valid is much smaller than the Hubble time, $\t\ll T_H$. Because of this limitation imposed on the time of applicability of the local frame, the local coordinates are also bounded in space by the radius, $r\ll R_H$, where $R_H=cT_H$ is the Hubble radius of the universe. The conclusion of this paragraph is that the LIC can be employed only for sufficiently close objects in the universe with the redshift factor $z\ll 1$ which excludes quasars and the most distant galaxies. Therefore, the formalism of the present paper is not applicable to the discussion of global cosmological properties and/or effects like the red shift of quasars.

The matching ensures that LIC can be constructed in the linearised Hubble approximation from the global coordinates, $y^\a$, by means of the special conformal transformation
(\ref{c4}) that respects EEP as the matching procedure demonstrates. In what follows, we accept the equalities, $\O(x(\eta))=a(\eta)$ and $a(\eta)=a(\t)$ that are valid in the linearised Hubble approximation. This is  because we work in the vicinity to the present epoch where $a(\t)=1+H\t+O(H^2\t^2)$ and $a(\eta)=1+{\cal H}\eta+O({\cal H}^2\eta^2)$ differ by the terms being quadratic with respect to the Hubble parameter which we neglect. It is worth noticing that in this approximation the special conformal transformation (\ref{c4}) coincides with the transformations to LIC having been found by other researchers who were trying to built local frames in cosmology with different mathematical techniques \citep{hongya:1924,2005ESASP.576..305K,2007CQGra..24.5031M}. The most important notice here is that the denominator of the coordinate transformation (\ref{c4}) is a non-linear function of the local coordinates $x^\a$ everywhere but on the light cone where it degenerates to the linear function because  the non-linear term, $x^2=\mm_{\a\b}x^\a x^\b=0$, vanishes on the light cone.

It should be emphasized that the tangent spacetime covered by LIC, $x^\a$, is locally diffeomorphic to the conformally-flat manifold \conf in the global coordinates $y^\a$ as they can be mapped onto each other by means of the coordinate transformation (\ref{c4}). Therefore, the tangent spacetime in LIC bears the same mathematical properties as \conf in the global coordinates $y^\a$. It is worth noticing that the tangent spacetime is considered, in accordance with EEP, as physical while \conf is not, in spite of their diffeomorphic equivalence. It brings in a certain conceptual difficulty in perceiving the physical meaning of the local time coordinate $x^0$ in equations of motion for light which can be misidentified with the proper time $\t$ of the Hubble observer while, in fact, $x^0\not=\t$ for photons. This misidentification occurs in the review paper \citep{2010RvMP...82..169C} and leads to erroneous conclusion on impossibility to observe the Hubble expansion locally. We shall explain and resolve this principal issue in next sections.

\section{The local inertial metric}\lab{sec4}

The local inertial coordinates, $x^\a$, are mathematical functions on FLRW manifold which have no immediate physical meaning unlike the Cartesian coordinates in Euclidean space. To make the local coordinates physically meaningful they should be further specified and operationally connected with measuring devices (clocks, rulers) of a set of some reference observers. The corresponding relations between the measuring tools and the local coordinates are known in differential geometry as inertial (or projective) structure \citep{eps_1972}. The Minkowski form of the physical local metric (\ref{act6}) suggests that LIC can be associated with the Gaussian normal coordinates based on the congruence of time-like geodesics of (electrically-neutral) test particles being at rest with respect to LIC \citep{waldorf}. 

The first step, is to identify the coordinate time $x^0$ with the proper time $\t$ of the Hubble observer at the origin of LIC. It can be done because the spacetime interval, $ds^2=-d(x^0)^2$ for $x^i=0$, and $ds^2=-d\t^2$ by the definition of the proper time \citep{waldorf}. The grid of the Gaussian coordinates start from the initial hypersurface, $\t=0$, that is orthogonal to the worldline of the Hubble observer. We identify the spatial Gaussian coordinates with the orthogonal (in the Euclidean sense) spatial coordinates $x^i$ of LIC on the initial hypersurface. Extension of the spatial coordinates from the initial hypersurface to arbitrary value of the time coordinate $x^0\equiv\t$ is performed by means of time-like geodesics. The Christoffel symbols of the local metric (\ref{act6}) are nil in a neighbourhood of the origin of LIC in accordance with the diffeomorphism (\ref{c4}) by which the LIC was introduced. More exactly, coordinate transformation (\ref{c4}) from the global to local coordinates, yields $\bar\G^{\a}_{\b\g}(x)=O(H^2x)$ that is beyond the linearized Hubble approximation under consideration. Because all the Christoffel symbols are nil, the time-like worldlines of particles having constant spatial coordinates, $x^i={\rm const.}$, are geodesics given by (\ref{bq1}). The proper time of the particle with the constant spatial coordinate $x^i$ coinsides with the time coordinate $x^0$ which was identified with the proper time of the Hubble observer. Hence, the parameter $\sigma$ in (\ref{bq1}) can be identified with the proper time $\t$ as well. After that equation (\ref{bq1}) describing the affine structure of the local Gaussian coordinates takes on the following simple form,  
\be\lab{bqq2}
\frac{d^2x^\a}{d\t^2}=0\;.
\ee 

The meaning of time-like geodesic equation (\ref{bqq2}) is as follows.
The world lines $x^\a=\le\{x^0=\t,x^i={\rm const.}\ri\}$ are identified with the network of static reference observers which play a fundamental role in local physical measurements. We admit that each static observer is equipped with an ideal (atomic) clock measuring their proper time which coincides with a time-like parameter, $x^0$, along the observer's worldline. Solving (\ref{bqq2}) reveals that $x^0\equiv\t$ is the proper time of the Hubble observer located at the origin of LIC. We assume that the ideal clocks of the static observers are synchronized. It can be done with Einstein's procedure of exchanging light signals as we will confirm in section \ref{bub3}. Eventually, the physical spacetime interval in the normal Gaussian coordinates reads
\be\lab{z11} 
ds^2=-d\t^2+\d_{ij}dx^idx^j\;.
\ee
It naturally coincides with the Minkowski spacetime interval in accordance with the formulation of EEP. 

It is worth emphasizing that the other Hubble observers who have constant value of the spatial global coordinates, $y^i={\rm const.}$, are moving with respect to the local inertial coordinates with a uniform velocity, $x^i=a(\t)y^i=(1+H\t)y^i$. It follows immediately from (\ref{c4}) after neglecting all terms being quadratic with respect to the Hubble constant $H$ and approximating $a(\eta)=1+H\eta=1+H\t=a(\t)$. In principle, the proper time $\t$ of the static observers is not exactly equal to the cosmological time $t$ of the Hubble observers since the latter move with respect to the former with the relative velocity $\beta^i=dx^i/d\t=Hy^i$ that is the velocity of the Hubble flow. Nonetheless, we can neglect the difference between $t$ and $\t$ because the two times are related by the special relativistic time transformation, $\t=t/\sqrt{1-\beta^2}\simeq t(1+\beta^2/2+...)$, and the small terms of the order of $\beta^2=O(H^2)$ can be neglected in the linearized Hubble approximation. Thence, we can interpret the local time $\t$ in the expression for local physical metric (\ref{z11}) as the cosmological time $t$ at any spatial point in the tangent space.

The Gaussian normal coordinates form a local inertial frame that is used for doing local physical measurements of time and space along time-like world lines of static observers and on space-like hypersurfaces of constant time. The frame is defined operationally in terms of the proper time of the ideal clocks and rigid rulers. The rulers are made of an ordinary matter which rigidity is defined primarily by the chemical bonds having an electromagnetic origin. We have proved \citep{Kopeikin_2012eph} that in the linearized Hubble approximation the electromagnetic (Coulomb) forces in an expanding universe remain the same as in a flat spacetime (see also section \ref{kmz3d}). For this reason, the rigid rulers and rods are not subject to the cosmological expansion and can serve for physical materialization of LIC. Another physical realization of the local Gaussian coordinates is achieved with the celestial ephemerides of the solar system bodies since their orbits are not affected by the Hubble expansion either \citep{2010RvMP...82..169C,Kopeikin_2012eph}.   

Any test particle that is in a free fall with respect to LIC moves along a geodesic with some affine parameter $\s$. However, we have to check whether the affine connection can be eliminated along the worldline of the particle so that the equation of the geodesic is reduced to (\ref{bq1}) or not. It can be definitely done along the worldlines of the static observers. However, due to the non-linear character of the coordinate transformation (\ref{c4}) not all components of the affine connection can be made nil on the light geodesics. It compels us to modify equation (\ref{bq1}) for photons to make it compatible with the conformal structure of the background cosmological spacetime. The modified equation is a light geodesic of a new {\it optical} metric introduced on the light cone which light rays are parametrized with the proper time $\t$ of the static observers. The reason for these complications is that the local time coordinate measured on the light cone, is a non-linear function, $x^0=x^0(\t)$, of the proper time $\t$ of the Hubble observer contrary to the erroneous result \citep{2010RvMP...82..169C} that the local time coordinate $x^0=\t$ for photons in cosmology.  The detailed form of function $x^0=x^0(\t)$ for photons is given in the next section.  

The concept of the optical metric was originally introduced by Gordon \citep{Gordon_1923} to study propagation of light in refractive medium. Later, it was elaborated on by Synge \citep{Synge_GRbook}, Ehlers \citep{Ehlers_1967} and Perlick \citep{perlick_2006,Perlick_2000}. More recent results can be found in \citep{2012PhRvD..86l4024N,PhysRevD.83.084047,Kantowski_2009} and references therein. It is worth mentioning that the concept of the optical metric has many similarities with the acoustic metric in analogue gravity theories \citep{lrr-2005-12}. 

\section{The local optical metric}\lab{opm123} 

The most precise measurements of spacetime events are made with electromagnetic waves and light \citep{2006LRR.....9....3W}. Therefore, we have to extrapolate the local metric (\ref{z11}) from the interior domain of the light cone on its hypersurface made of the light-ray geodesics. The metric on the light cone hypersurface is called the optical metric that is a tool for investigating motion of freely-propagating photons \citep{Synge_GRbook}. The extrapolation is straightforward in the case of Minkowski spacetime that is tangent to a pseudo-Riemannian manifold which is not conformally-flat. For such a manifold the metric (\ref{z11}) would be valid everywhere including the light cone hypersurface. However, FLRW spacetime manifold is conformally-flat and the construction of the local inertial coordinates is fulfilled with the help of the special conformal transformation (\ref{c4}) which work differently for time-like and light-like intervals.  

The reason for this difference is that by doing transformation (\ref{c4}) we build a physical tangent spacetime to FLRW manifold which has the flat metric $\mm_{\a\b}$ with the affine connection being nil. Therefore, the affine structure of the tangent spacetime \tang is locally diffeomorphic to the conformally-flat manifold \conf. Light geodesics in \conf obey to differential equation 
\be\lab{ce5}
\frac{d^2y^\a}{d\eta^2}=0\;,
\ee 
which is uniformly parametrized with the conformal time $\eta$ \citep{waldorf}.
If we approved EEP and assumed that the affine connection is nil everywhere including the light cone it would allow us to write light geodesics in the form of equation (\ref{bq1a}) which is parametrized with the proper time $\t$ of the Hubble observer. Comparing (\ref{bq1a}) with (\ref{ce5}) would force us to admit that the proper time $\t$ is equal to the conformal time $\eta$ up to a scale (conformal) factor, $a(\eta)$, of FLRW universe. However it would violate the identity of the proper time $\t$ of the Hubble observer with the cosmological time $t$ which is a cornerstone of modern cosmology being confirmed by cosmological observations of various type \citep{2005pfc..book.....M,weinberg_2008}. This dichotomy of the proper time $\t$ is unacceptable and can be resolved if we shall abandon EEP for photons and treat the local metric $g_{\a\b}$ on the light cone differently from the flat metric $\mm_{\a\b}$. This brings in the concept of the optical metric.

To elaborate on this issue, let us assume for a while that the flat inertial metric (\ref{z11}) is valid on the light cone. Then, light moves along null geodesics defined locally by equation (\ref{bq1a}) where $\t$ is the proper time of the static observer. A solution of this equation is a linear function of time,  $x^0=\t,\;x^i=k^i\t$ where $k^i$ is a unit vector in the direction of propagation, $\d_{ij}k^ik^j=1$. This local solution must be compatible with that obtained from the equation of light geodesics (\ref{ce5}) derived from the FLRW metric (\ref{q1}) in the global conformal coordinates, that is $y^0=\eta,\; y^i=k^i\eta$. 

The local coordinates, $x^\a$, and the global conformal coordinates, $y^\a$, of the photon are related by the conformal transformation (\ref{c4}) which should be taken on the light cone where, $x^2=\mm_{\a\b}x^\a x^\b=0$. Under the assumptions we have made, it reads
\ba\lab{si1}
\t&=&a(\eta)\eta\;,\\
\lab{zx1}
x^i&=&a(\eta)y^i\;,
\ea
and is consistent with the solutions of the light geodesics in the global and local coordinates. Indeed, after substituting $x^i=k^i\t$ and $y^i=k^i\eta$ to (\ref{zx1}) we get (\ref{si1}).  

So far, everything seems to be right but the reader is to remember that according to the operational construction of the Gaussian coordinates the proper time of the static observers, $\t$, coincides with the cosmological time $t$, that is $\t\equiv t$. Hence, if the affine connection also vanishes on the light cone, equation (\ref{si1}) must yield the following relation between the cosmological and conformal times, 
\be\lab{si2} 
t=a(\eta)\eta=\eta+H\eta^2+O(H^2)\;.
\ee
However, this equation contradicts the original definition (\ref{oka1}) of the conformal time $\eta$ in cosmology which, in the linearised Hubble approximation, reads
\be\lab{meo8} 
t=\int_0^\eta a(\eta')d\eta'=\eta+\frac{H}2\eta^2+O(H^2)\;.
\ee
Apparently, equation (\ref{si2}) disagrees with (\ref{meo8}). This contradiction points out that the affine connection cannot be eliminated on the light cone in the local Gaussian coordinates. This pitfall was overlooked in \citep{2010RvMP...82..169C}. 

Let us now establish the optical metric on the light cone. To avoid possible notational confusion we shall denote the time coordinate of LIC measured on the light cone as $\l$ and look for the functional dependence $\l=\l(\t)$. Let a photon be emitted from the origin of LIC at instant $\t=\l=0$ and propagate in the direction given by a unit vector $k^i$. We apply the coordinate transformation (\ref{c4}) to compare the values of $\t$ taken on the worldline of the Hubble observer and $\l$ taken on the light ray. Spatial coordinates of the origin of LIC, $x^i=0$, while the spatial coordinate of light-ray is approximately $x^i=k^i\t$. It gives us $x^2\equiv\mm_{\a\b}x^\a x^\b=-\t^2$ on the worldline of the origin of LIC, while $x^2=0$ on the light-ray geodesic. Substituting these values for $x^2$ into the conformal transformation (\ref{c4}) yields 
\bsu\ba\lab{oo8a}
a(\eta)\eta&=&\t+\displaystyle\frac{H}{2}\t^2\qquad\mbox{\small({\it the Hubble observer})},\\\lab{oo8b}
a(\eta)\eta&=&\lambda\phantom{+\displaystyle\frac{H\t^2}{2\gamma^2}}\qquad\mbox{\small({\it the light ray})}.
\ea\esu
Notice that (\ref{oo8b}) replaces (\ref{si1}) that was proven to be erroneous. 

The important point to understand is that the numerical value of the global time coordinate $\eta$ for the Hubble observer is the same as that for the propagating photon because both, the Hubble observer and photons obey equation (\ref{ce5}) in the global coordinates which solution is $y^0=\eta,\;y^i=0$ for the Hubble observer and $y^0=\eta,\;y^i=k^i\eta$ for photons. Hence, the numerical values of function $a(\eta)\eta$ in the left
side of (\ref{oo8a}) and (\ref{oo8b}) are equal. It gives us the transformation between the time coordinate $\t$ measured by the Hubble observer and the time coordinate $\l$ measured on the light ray, which we were looking for. More specifically, 
\be\lab{nbr5}
\lambda=\t+\displaystyle\frac{H}{2}\t^2\;,
\ee or in a differential form
\be\lab{of1}
d\lambda=a(\t)d\t\;,
\ee
where we have made use of the linearised Hubble approximation of the scale factor $a(\t)=1+H\t$. Apparently, function $\l$ is not a linear function of the proper time $\t$ which is a clear indication that the metric on the light cone is different from the Minkowski metric.

The optical metric, $g_{\a\b}$, is defined in terms of the differentials of the time coordinate $\lambda$ on the light geodesics and that of the spatial coordinates $x^i$,
\be\lab{wky7}
ds^2=-d\l^2+\d_{ij}dx^i dx^j\;.
\ee 
It looks similar to the local inertial metric (\ref{z11}) with the time coordinate $\t$ replaced with time $\lambda$.
Substituting (\ref{of1}) into (\ref{wky7}) establishes the optical metric in terms of the Gaussian normal coordinates $x^\a=(\t,x^i)$,
\be\lab{fe3}
d s^2=-a^2(\t)d\t^2+\d_{ij}dx^i dx^j\;,
\ee
or $g_{\a\b}={\rm diag}[-a^2,1,1,1]$.
The contravariant components $g^{a\b}$ of the optical metric are defined by equation $g_{\a\b}g^{\b\g}=\d_a^\g$ which yields $g^{\a\b}={\rm diag}[-a^{-2},1,1,1]$. 
The optical metric coincides with the Minkowski metric at the initial instant of time $\t=0$ but its time-time component deviates from it linearly as time goes on. 

The optical metric must be used for rising and lowering indices of the wave co-vector of electromagnetic wave, which is defined by $k_\a=\pd_\a\varphi$, where $\varphi$ is the phase of electromagnetic wave. The reason behind this rule has been explained by Synge \citep{Synge_GRbook} -- the contravariant optical metric, $g^{\a\b}$ contracted with $k_\a k_\b$ serves as a Hamiltonian function $H=g^{\a\b}k_\a k_\b$ in the description of motion of photons. The Hamilton-Jacobi equations of motion for photons tells us immediately that it is the optical metric which must be used to rise and lower the indices of the wave co-vector. We conclude that the wave vector $k^\a=dx^\a/d\s$ where $\s$ is the canonical parameter along the light geodesic is related to $k_\a$ by the following rule $k^\a=g^{\a\b}k_\b$ or $k_\a=g_{\a\b}k^b$.
  
Equation (\ref{fe3}) formally includes terms of the second order $O(H^2)$ with respect to the Hubble constant, but all such terms must be neglected in accordance with the linearized Hubble approximation adopted in the present paper, for example,  $a^2(\tau)=1+2H\tau$, and so on. 
Notice that the optical metric (\ref{fe3}) is degenerated in the sense that it is valid only on the light cone hypersurface. It serves for describing propagation of freely falling photons in the local Gaussian coordinates but cannot be applied to events separated by time-like or space-like intervals to which the inertial metric (\ref{z11}) must be applied. 

\section{The light geodesics}\lab{lglc}

The optical metric tells us that on the hypersurface of light ray geodesics, the Christoffel symbols of the metric are, $\G^{\a}_{00}=Hu^\a$, where $u^\a=(1,0,0,0)$ is a four-velocity of the static observers in LIC, and all other Christoffel symbols vanish. These values of the Christoffel symbols can be also obtained by direct application of the time transformation (\ref{nbr5}) to the Christoffel symbols $\bar\G^{\a}_{\b\g}(x)=$ in the Gaussian coordinates $x^\a=(\t,x^i)$.

We define the wave vector of propagating electromagnetic wave as $k^\a=dx^\a/d\s$ where $\s$ is an affine parameter along the light ray. Equation of the parallel transport of the wave vector
\be\lab{wkuku}
\frac{dk^\a}{d\s}+\G^\a_{\b\g}k^\b k^\g=0\;.
\ee
After replacing the Christoffel symbols $\G^\a_{\b\g}$ in (\ref{wkuku}) with their values calculated above from the optical metric we get the equation of the parallel transport of the wave vector 
\be\lab{ert2}
\frac{dk^\a}{d\s}+H u^\a=0\;,
\ee 
where we have approximated the time component of the wave vector in the term being proportional to the Hubble constant as $k^0=1+O(H)$ to simplify the equation.

Equation (\ref{ert2}) is equivalent to the geodesic equation for coordinates of the light ray \citep{kopeikin_2013}, 
\be\lab{wku1}
\frac{d^2x^\a}{d\s^2}+Hu^\a=0\;.
\ee
It differs from equation of motion (\ref{bq1}) which might be expected on the ground of applicability of EEP to photons that assumes that light rays in vacuum move locally along geodesics of the flat metric $\mm_{\a\b}$. The difference emerges due to the presence of the time-dependent conformal factor in the equations establishing the equivalence between the points of FLRW manifold and its tangent spacetime.  

Replacing the affine parameter $\s$ with the proper time $\t$ of the static observers brings the geodesic equation (\ref{wku1}) to
\be\lab{mdu5}
\frac{d^2x^\a}{d\t^2}=H\left(\frac{dx^\a}{d\t}-u^\a\ri)\;.
\ee
This equation agrees with the derivation given in our paper \citep{Kopeikin_2012eph}. It also agrees with the light geodesic equation (\ref{ce5}) derived in the global conformal coordinates $y^\a$ of FLRW metric. Solution of (\ref{mdu5}) for radial geodesics is given in section \ref{d4d}. 

Equation (\ref{mdu5}) predicts the existence of an ``anomalous'' force exerted on a freely-falling photon as contrasted with EEP prediction and directed toward the origin of LIC for outgoing light ray It should not be misinterpreted as a violation of general relativity or Newtonian gravity like the ``fifth force'' \citep{1992Natur.356..207F}. In fact, equation (\ref{mdu5}) is a direct consequence of general relativity and the conformal invariance of light ray geodesics applied along with the cosmological principle stating that the global cosmological time $t$ must be identical with the proper time $\t$ measured by the Hubble observers. It explains how the Hubble expansion of the universe may appear locally. We discuss the observational aspects of this local cosmological effect in next section in more detail.

\section{Cosmological effects in the local frame}\lab{rr1}

\subsection{Physical realization of the local Gaussian coordinates}
We shall use the normal Gaussian coordinates, $x^\a$, along with the optical metric (\ref{fe3}) to explore the fundamental physics of the Hubble expansion in the solar system. It requires a practical realization of the
normal Gaussian coordinates that is an identification of spacetime points with the set of static observers who are not subject to the Hubble expansion locally. The fact that such observers physically exist was
established by many researchers \citep{1999CQGra..16.1313B,2007CQGra..24.5031M,2006PhRvD..74f4019C,2010RvMP...82..169C,Kopeikin_2012eph}. who proved that both the Coulomb electric force and the Newtonian gravitational force are not affected locally by the Hubble expansion of the background FLRW spacetime, at least, in the linearised Hubble approximation. It means that
we can physically manufacture rigid rods and rulers which length is not subject to the cosmological influence. The only problem remains is to control and subtract the environmental thermal and mechanical deformations that can be solved in a standard way, for example, by making use of temperature-controlled laser metrology \citep{2009homp.book..365S}. A set of such rigid rods materializes the orthogonal grid of the normal Gaussian coordinates and represent the reference static observers in laboratory. 

Planetary orbits in the solar system (including the
orbits of space probes) obey the Newtonian equations of motion with the cosmological tidal terms that are of the order of $O(H^2)$, and are negligibly small \citep{2007CQGra..24.5031M}. Therefore, celestial ephemerides of the solar system bodies provide a dynamic approach to the physical realization of the normal Gaussian coordinates in the solar system \citep{Kopeikin_2012eph} by mapping planetary positions and velocities on the initial hypersurface of constant barycentric time with the (post-Newtonian) equations of motion \citep{kopeikin_2011book,2013strs.book.....S}. For this reason, the IAU fundamental ephemerides are a powerful instrument for doing navigation of spacecrafts in deep space and a unique research tool of relativistic gravitational physics \cite{iau:2012}. 

The rest of this section describes several physical experiments to measure the Hubble expansion locally in the solar system without invoking observations of distant quasars and/or cosmic microwave background radiation (CMBR).

\subsection{Radar and laser ranging}\lab{d4d}
Precise dynamical modelling of orbital and rotational motion of astronomical bodies in the solar system (major and minor planets, asteroids, spacecraft, etc..) is inconceivable without radar and laser ranging. The ranging is an integral part of the experimental testing of general relativity and alternative theories of gravity in the solar system \citep{2006LRR.....9....3W,kopeikin_2011book,Williams_2012}. We are to check if the Hubble expansion can be measured in the ranging experiments.

Equation of light propagation in the local Gaussian coordinates $x^\a=\le(\t,x^i\ri)$ is given by the light cone equation, $ds^2=0$, with the optical metric (\ref{fe3}) which explicit form reads
\be\lab{fe4}
\d_{ij}dx^i dx^j=a^2(\t)d\t^2\;.
\ee
Let us consider a radial propagation of light from (or to) the origin of the local coordinates where the central (Hubble) observer is located. The radial (always positive) spatial coordinate of a photon is, $r=\sqrt{\d_{ij}x^i x^j}$, and the equation of light's radial propagation (\ref{fe4}) implies
\be\lab{fe5}
dr=\pm a(\t)d\t\;,
\ee
where the sign plus/minus corresponds to the outgoing/incoming light ray. 
The reader may notice that the coordinate speed of light, $|dr/d\t|=a(\t)$, can exceed the fundamental value of $c=1$. There is no violation of special relativity here because this effect is non-local. The local value of the speed of light measured at time $\t$, is always equal to $c=1$ after a corresponding re-scaling of units to make the scale factor $a(\t)=1$. This is also true for any other local measurement done at a different point in space because the group of the conformal isometry includes the Poincare group as a sub-group which allows us to change the initial epoch and the position on the background manifold without changing the differential equation (\ref{fe5}). 

Let a light pulse be emitted at time $\t_0$ at point $r_0$, riches the target at radial coordinate $r>r_0$ at time $\t$, and returns to the point of observation at radial distance $r_1<r$ at time $\t_1$. Equations of propagation of the outgoing and incoming light rays are obtained from (\ref{fe5}), and read
\be\lab{fr4r}
r=r_0+\int^\t_{\t_0} a(\t')d\t'\;,\qquad r_1=r-\int_\t^{\t_1} a(\t')d\t'\;.
\ee
The result of the integration is fully determined by the boundary conditions imposed on the solution. More specifically, we demand that at the time of emission, $\t_0$, the coordinate speed of light $\dot r(\t_0)=1$ which means that $a(\t_0)=1$. For any other instant of time the Taylor expansion of the scale factor yields
\be\lab{kib5}
a(\t)=1+H\left(\t-\t_0\ri)+O\left(H^2\ri)\;.
\ee
Replacing the scale factor $a(\t')$ in the integrands of integrals in (\ref{kib5}), and integrating brings about the following results
\begin{subequations}
\ba\lab{tt8a}
r&=&r_0+(\t-\t_0)+\frac{H}2\left(\t-\t_0\ri)^2\;,\qquad\qquad\qquad\qquad\;\mbox{(\it outgoing ray}),\\\lab{tt8b}
r_1&=&r\phantom{_0}-(\t_1-\t)-\frac{H}2 \left[(\t_1-\t_0)^2-(\t-\t_0)^2\right]\;,\qquad\mbox{(\it incoming ray}).
\ea
\end{subequations}

Let us assume for simplicity that the radar ranging is conducted by the Hubble observer at the origin of the local coordinates so that both the points of emission and observation of the light signal are at the origin and have the radial coordinate, $r_0=r_1=0$. 
We define the {\it radar distance} by a standard equation \citep{LanLif,kopeikin_2011book}
\be\lab{tt9}
\ell\equiv\frac12\le(\t_1-\t_0\ri)\;,
\ee
which is a relativistic invariant due to the covariant nature of the proper time $\t$ and the constancy of the fundamental speed $c=1$ in the geometrized system of units adopted in the present paper. After solving (\ref{tt8a}), (\ref{tt8b}) we obtain 
\ba\lab{tt9a}
\t&=&\frac12\left(\t_0+\t_1\ri)+\frac12Hr^2\;,\\
\nonumber\\
\lab{tzi9}
\ell&=&r-Hr^2\;,
\ea
where the residual terms of the order of $O(H^2)$ have been neglected, $r$ is the radial distance of the point of reflection of a radar signal at time $\t$, and the value of the scale factor $a(\t)$ is taken at the time of reflection of the radar signal as well. 

This calculation reveals that the difference between the coordinate distance $r$ and the invariant {\it radar distance} $\ell$ is of the order of $Hr^2$. Planetary ranging is done for the inner planets of the solar system so we can approximate $r\simeq 1$ astronomical unit (au) and $H\simeq 2.3\times 10^{-18}$ s$^{-1}$. Hence the difference $Hr^2\simeq 0.17$ mm which is a factor of $\sim 10^4$ smaller than the current ranging accuracy ($\sim 2$ m) to interplanetary spacecraft \citep{Folkner_2009IPNPR,Fienga_2009AA}. In case of lunar laser ranging to the Moon, the coordinate radius of the lunar orbit $r\simeq 384,000$ km, and the estimate of the residual term $Hr^2\simeq 1.1\times 10^{-7}$ sm which is one million times less than the current accuracy ($\sim$ 1 mm) of LLR \citep{murthyetal08}. We conclude that in radar/laser ranging experiments:
\begin{enumerate}  
\item[(1)] within the measuring uncertainty the coordinate radial distance $r=\ell$,
\item[(2)] the radial distance $r$ in the local frame of reference has an invariant geometric meaning in agreement with the definition of the proper distance accepted in cosmology \citep{weinberg_1972,mukh_book},
\item[(3)] the radar/laser ranging metrology is insensitive to the Hubble expansion in the local coordinates.  
\end{enumerate} 
Hence, the celestial ephemerides of the solar system bodies built on the basis of radar/laser ranging data are not crippled by the Hubble expansion. They represent a dynamical reference frame with a fixed value of the astronomical unit (au) which is not changing in time and can be treated as a rigid ruler for measuring distances between celestial bodies within the solar system in accordance with a recent resolution of IAU General Assembly (Beijing 2012) on the meaning and value of astronomical unit \citep{2012IAUJD...7E..40C}.

\subsection{Einstein's synchronization of clocks}\lab{bub3}
Let us now consider the Einstein procedure of the synchronization of two clocks based on exchange of light signals between the clocks. We want to synchronize the clock of the central Hubble observer with the clock of a static observer located at a point with the Gaussian radial coordinate $r$. We apply exactly the same procedure as in the case of radar ranging described above. By Einstein's definition, when the photon riches the reflection point with the radial coordinate $r$ at the instant of time $\t$, the clock of the Hubble observer at the point, $r=0$, reads the time
\be\lab{tt10}
\t^*=\frac12\le(\t_0+\t_1\ri)\;,
\ee
because the time rate of the (ideal) clock of the Hubble observer is uniform. The instant of time $\t^*$ is defined as being simultaneous with the time reading, $\t$, of a second clock located at the position with a radial coordinate, $r$, at the instant when the light signal is reflected. The time $\t^*$ as a function of $\t$, can be found immediately from (\ref{tt9a})
\be\lab{tt10a}
\t^*=\t-\frac12Hr^2\;.
\ee
This relation reveals that in order to synchronize two clocks separated by a radial distance $r$, we have to subtract the time difference $Hr^2$ from the reading $\t$ of the clock of the static observer at the point with radial coordinate $r$ in order to make the time readings of the two clocks identical. Because the radial distance $r$ coincides with the invariant radar distance $\ell$, which is a measurable quantity, the Einstein synchronization of clocks in such experiment is operationally possible.

The two clocks will remain synchronized as time goes on, if and only if, the radial distance between the clocks does not change. For example,  a clock at a geocenter will remain synchronized with clocks on-board of a geostationary satellite moving around Earth on a circular orbit. On the other hand, an ultra-stable clock on board of spacecraft which moves with respect to the primary time standard on Earth may detect the de-synchronization effect due to the Hubble expansion of the universe if the radial distance between Earth and the spacecraft changes periodically. If the change in the radial distance amounts to $\d r$, the overall periodic time difference caused by the clock's de-synchronization amounts to $\d\t=\d(\t^*-\t)=2H(r/c)^2(\d r/r)$. Expressing $r$ in astronomical units we can find a numerical estimate of the de-synchronization between the readings of the two clocks, 
\be\lab{tt10b}
\d\t=1.7\times 10^{-12}\le(\frac{r}{1\;{\rm au}}\ri)^2\left(\frac{\d r}{r}\ri) \;\mbox{[s]}\;,
\ee
where we have used the approximate numerical value of the Hubble constant, $H=2.3\times 10^{-18}$ s$^{-1}$, the universal speed $c=3\times 10^{10}$ cm/s in cgs units, and the astronomical unit (au) $=1.5\times 10^{13}$ cm. This local cosmological effect may be detectable by NIST and/or other world-leading timekeepers.
\subsection{Doppler effect in the local frame}\lab{glb}
\subsubsection{Theory of the Doppler effect}

Next step is to consider the Doppler effect that is a change in frequency of propagating electromagnetic wave (light) emitted at one spacetime event and received at another one, as caused by various physical reasons - relative motion of observer and the source of light, gravity field, expansion of the Universe, etc.. A monochromatic electromagnetic wave propagates on a light cone hypersurface of a constant phase $\varphi$, that is a function of spacetime coordinates, $\varphi=\varphi(x^\a)$. The wave (co)vector is $k_\a=\pd_\a\varphi$, and frequency of the wave measured by an observer moving with 4-velocity, $u^\a$, is \citep{waldorf,kopeikin_2011book}
\be\lab{d1}
\omega=-k_\a u^\a\;.
\ee
The definition of frequency of electromagnetic wave does not depend on the metric tensor and can be calculated directly as soon as we know $k_\a$ and $u^\a=dx^\a/d\t$. Indeed, 
\be\lab{n4d5}
\omega=-k_a u^\a=\frac{\pd\varphi}{\pd x^\a}\frac{dx^a}{\d\t}=\frac{d\varphi}{d\t}\;,
\ee
which is just the rate of change of the phase of electromagnetic wave along the world line of observer.
 
Let us denote the point of emission of the wave $P_1$, the point of its observation $P_2$, and the emitted and observed wave frequencies as $\omega_1$ and $\omega_2$, respectively. Their ratio
\be\lab{d2}
\frac{\omega_2}{\omega_1}=\frac{\le(k_\a u^\a\ri)_{P_2}}{\le(k_\a u^\a\ri)_{P_1}}\;,
\ee
quantifies the Doppler effect. The present paper is not concerned with special relativistic Doppler effect due to the motion of emitter and/or observer (see \citep{kopeikin_2011book,2002PhRvD..65f4025K} for further mathematical details) but focuses on the cosmological frequency shift measured by static observers in the local inertial frame in the presence of the cosmological expansion.
Locally, four-velocity of the static observers, $u^\a=(1,0,0,0)$. Hence, formula (\ref{d2}) reads
\be\lab{dd2q}
\frac{\omega_2}{\omega_1}=\frac{k_0(\t_2)}{k_0(\t_1)}\;,
\ee
where $\t_1$ and $\t_2$ are the instants of emission and observation of light respectively. 

It is worth recalling that co-vector $k_\a$ and vector $k^\a$ are both lying on the light cone hypersurface. Hence, they are connected by the optical metric $g_{\a\b}$ that is $k_a=g_{\a\b}k^\a$, and $k^\a=g^{\a\b}k_\b$ where the optical metric satisfies to $g_{\a\b}g^{\b\g}=\d_\a^\g$, where $\d_\a^\g$ is the Kroneker symbol. For this reason, the frequency of light $\omega=k_\a u^\a=g_{\a\b}k^\b u^\a$. In other words, it is the optical metric, $g_{\a\b}$, which is used in order to define the frequency of light in terms of the scalar product between the wave vector, $k^\a$ and the four-velocity of observer $u^\a$.  We emphasize that only equation (\ref{d1}) is a unique and unambiguous definition of electromagnetic frequency, $\omega$. The principles of mathematical operations with the optical metric and calculation of observables in the present paper are identical to those applied in the optics of dispersive media \citep{Synge_GRbook,Ehlers_1967,perlick_2006,Perlick_2000}. 

The parallel transport of $k_0$ along the light geodesic must be computed with the optical metric (\ref{fe3}). Equation of the parallel transport of $k_\a$ in the local coordinates can be obtained from (\ref{wkuku}). It reads,
\be\lab{po1}
\frac{dk_\a}{d\s}=-\frac12 \frac{\pd{g}^{\m\n}}{\pd x^\a}k_\m k_\n\;,
\ee
where $\s$ is the affine parameter along the light ray, the wave vector of light $k^\a=dx^\a/d\s$, $k_\a={g}_{\a\b}k^\b$, the optical metric ${g}_{\a\b}={\rm diag}\le[-a^2,1,1,1\ri]$, ${g}^{a\b}={\rm diag}\le[-a^{-2},1,1,1\ri]$, and the scale factor $a\equiv a(\t)$. 

The optical metric depends on time $\t$ only. Therefore, the spatial component of equation (\ref{po1}) immediately tells us that $k_i$ and, hence, $k^i=\d^{ij}k_j$ remain constant along the light ray. The time component of (\ref{po1}) reads
\be\lab{fgt3}
\frac{dk_0}{d\s}=\frac12\frac{d}{d\t}\le(\frac1{a^2}\ri)k^2_0\;.
\ee
The left side of (\ref{fgt3}) can be recast to 
\be\lab{po2}
\frac{dk_0}{d\s}=k^\b\frac{\pd k_0}{\pd x^\b}=k^0\frac{dk_0}{d\t}=-\frac{k_0}{a^2}\frac{dk_0}{d\t}\;,
\ee
where we have made use of the fact that $k^0=g^{0\a}k_\a=g^{00}k_0=-k_0/a^2$. 
Replacing (\ref{po2}) in (\ref{fgt3}) brings it to the following form
\be\lab{d3b}
\frac{dk_0}{d\t}=-\frac{a^2}2\frac{d}{d\t}\le(\frac1{a^2}\ri)k_0\;.
\ee
Solution of this equation is
\be\lab{d3c}
\frac{k_0(\t)}{a(\t)}={\rm const.}
\ee
and we use it in (\ref{dd2q}). It yields the Doppler shift of frequency of electromagnetic wave in the expanding universe as measured in the local inertial frame 
\be\lab{d5a}
\frac{\omega_2}{\omega_1}=\frac{a(\t_2)}{a(\t_1)}=1+H(\t_2-\t_1)\;,
\ee
and it does not depend on the choice of the epoch $\t_{\rm e}$ when $a(\t_{\rm e})=1$. For this reason, the Doppler shift equation (\ref{d5a}) has a universal character. Moreover, equation (\ref{d5a}) tells us that the cosmological Doppler shift measured by the local static observers is {\it blue} because $\t_2>\t_1$ and $a(\t_2)>a(\t_1)$ due to the Hubble expansion and, consequently, $\o_2>\o_1$. 

On the other hand, we know that the
cosmological Doppler shift of frequency of light of distant quasars is {\it red} \citep{weinberg_2008}. There is no contradiction here with our result (\ref{d5a}). Cosmological red shift is measured with respect to the reference objects (quasars)
which have fixed values of the global coordinates, $y^i$, while the local Doppler shift (\ref{d5a}) is measured with respect to static observers having fixed Gaussian coordinates $x^i$. Thus,
the Doppler shift measurements in the global cosmological spacetime and in the local tangent spacetime refer to two different sets of observers moving one with respect to another with the velocity of the Hubble flow.
Therefore, it is natural to expect a different signature of the Doppler effect -- {\it red} shift for light coming from distant quasars and {\it blue} shift for light emitted by the astronomical objects, for example spacecraft, within the solar system. Our theory provides an exact answer for the signature and magnitude of the cosmological {\it blue} shift effect measured in the local inertial frame. 

The Doppler effect in the tangent spacetime of FLRW universe has been considered by a number of other authors, most notably by Carerra and Giulini \citep{2006CQGra..23.7483C,2010RvMP...82..169C}. They claimed that the cosmological expansion does not produce any noticeable Doppler effect in the local radio-wave frequency measurements. Their conclusion is invalid as they erroneously identified the local time coordinates $x^0$ with the proper time $\t$ of the Hubble observer on a worldline of any freely-moving particle including photons. We discussed in section \ref{opm123} the reasons why such an identification is misleading and leads to a wrong result for the local Doppler shift. Hence, the arguments by Carerra and Guilini against observability of cosmological expansion in the local frames on FLRW manifold are misleading and inapplicable for discussing gravitational physics of photons in the local frame.

\subsubsection{Measuring the Hubble constant with spacecraft Doppler-tracking}\lab{mdt5nc}
Precise and long-term Doppler tracking of space probes in the solar system may be a new way to measure the local value of the Hubble constant $H$. It is highly plausible that the ``Pioneer anomaly'' detected by John Anderson \citep{2002PhRvD..65h2004A} with the JPL deep-space Doppler tracking technique in the hyperbolic orbital motion of Pioneer spacecraft, is caused by the Hubble expansion which changes the frequency of radio waves in the spacecraft radio communication link in an amazing agreement (both in sign and in magnitude) with our equation (\ref{d5a}). 

More detailed theory supporting the cosmological origin of the ``Pioneer anomaly'' effect is given in our paper \citep{Kopeikin_2012eph} where we have derived the local equations of motion for charged and neutral test particles as well as for photons, in the FLRW universe. We have proved that in the local frame of reference the equations of motion for interacting massive neutral and/or charged particles do not include the linear terms of the first order in the Hubble constant. On the other hand, equations of motion of photons do contain such linear terms of the order of $H$ which have dimension of acceleration. 

The present paper confirms our previous result from the point of view of a set of local observers doing their measurements in the tangent space of the FLRW manifold. Transformation to the local coordinates $x^\a=(x^0,x^i)$ allows us to transform FLRW metric to the Minkowski metric $ds^2=-(dx^0)^2+\d_{ij}dx^idx^j$ but the time coordinate $x^0$ considered as a function of the proper time $\t$ of the Hubble observer, $x^0=x^0(\t)$, has different mathematical values on worldlines of static observers and on those of photons. It inevitably introduces two local metrics in the tangent spacetime. The inertial Minkowski metric, $\mm_{\a\b}$, serves for physical measurements on static or slowly-moving objects while the optical metric, $g_{\a\b}$, describes the local geometry on light-like hypersurface as observed by the static observers. The existnece of the two metrics in the local frame explains why the Pioneer anomalous acceleration is suppressed for planets but is present in the motion of light. The optical metric leads to a non-vanishing affine connection on the light cone which can be formally interpreted as a violation of EEP for photons. The EEP violation manifests in observation of the Pioneer anomalous acceleration in John Anderson's experiment \citep{2002PhRvD..65h2004A}. 

Equation (\ref{d5a}) explains the ``Pioneer anomaly'' effect as a consequence of the {\it blue} frequency shift of radio waves on their round trip from Earth to spacecraft and back. Let us denote the emitted frequency $\o_0$, the frequency received at spacecraft and transmitted back to Earth $\o_1$, and the frequency received on Earth $\o_2$. Then, according to (\ref{d5a}), the shift between $\o_0$ and $\o_2$ is
\be\lab{fgl9}
\frac{\o_2}{\o_0}=\frac{\o_2}{\o_1}\frac{\o_1}{\o_0}=\frac{a(\t_2)}{a(\t_1)}\frac{a(\t_1)}{a(\t_0)}=\frac{a(\t_2)}{a(\t_0)}= 1+H(\t_2-\t_0)\;,
\ee
where $\t_0$ is the time of emission on Earth, $\t_1$ is the time of receiving and re-transmission of the signal at spacecraft, and $\t_2$ is the time of reception of the signal on Earth.
The Doppler shift is defined as $z\equiv (\o_2/\o_0)-1$, and we have to add the special relativistic term $-v/c$ due to the outward radial motion of Pioneer spacecraft with radial velocity $v=v(\t_1)$ (we prefer to keep here the speed of light $c$ explicitly). We get 
\be\lab{sxw2}
z=-\frac{v}c+H(\t_2-\t_0)\;,
\ee
that shows that the cosmological shift of frequency appears as a tiny {\it blue} shift on top of much larger {\it red} shift of frequency caused by the outward motion of the spacecraft. It fully agrees with the observations \citep{2002PhRvD..65h2004A}.

The time rate of change of the Doppler shift is $\dot z\equiv dz/d\t_2$ which yields
\be\lab{srb6}
\dot z=\frac{a}{c}+H=\frac1c\le(a+Hc\ri)\;,
\ee
where $a=-dv/d\t_1$ is a deceleration of the Pioneer spacecraft due to the attraction of the solar gravity field, and we have neglected all terms of the order of $1/c^2$ so that, for example, $dv/d\t_2=dv/d\t_1$. The deceleration, $a$ is positive because the derivative, $dv/d\t$, is negative due to the gravitational pull of the Sun. Thus, the Hubble frequency-shift term, $Hc$, is added to the positive {\it deceleration} and can be interpreted as a constant, directed-inward acceleration, $a_P=Hc$, in the motion of spacecraft. In fact, such an interpretation is misleading as the true cause of the ``anomalous'' acceleration is associated with the motion of photons but not the spacecraft. This is the reason why the vigorous attempts to find out the explanation for the ``anomalous gravity force'' exerted on Pioneer spacecraft were unsuccessful. The observed value of $a^{\rm obs}_P=8.5\times 10^{-10}$ m$\cdot$s$^{-2}$ \citep{2002PhRvD..65h2004A} is in a good agreement (both in sign and in magnitude) with the theoretical value of $a^{\rm theory}_P=Hc\simeq 7\times 10^{-10}$ m$\cdot$s$^{-2}$. Our theoretical result provides a strong support in favour of general-relativistic explanation of the ``Pioneer anomaly'' as caused by the conformal geometric structure of FLRW cosmological manifold.  

The observed value of $a^{\rm obs}_P$ exceeds theoretical value $a^{\rm theory}_P$ by 20\%. This can be explained as the recoil of the thermal emission of electric power generators (RTG) on board of the spacecraft \citep{2002PhRvD..65h2004A,2007AdSpR..39..291T}. Indeed, recent studies \citep{2011AnP...523..439R,Turyshev_2012} indicate that the numerical value of the Pioneer anomalous acceleration may be slightly decreasing over time which may indicate to the radioactive decay of the power generators of Pioneer spacecraft. This observation supports the idea that thermal emission is partially responsible for the Pioneer effect. The question is how much it contributes to the overall effect. The papers  \citep{2011AnP...523..439R,Turyshev_2012} state the Pioneer effect is 100\% thermal. We do not think so. 

First of all, we notice that the geometric theory of the Pioneer effect having been worked out in \citep{Kopeikin_2012eph} and in the present paper points out that the numerical value of the Pioneer effect cannot be smaller than $a^{\rm theory}_P=7\times 10^{-10}$ m$\cdot$s$^{-2}$. The thermal emission always adds to the general-relativistic theoretical prediction, $a^{\rm theory}_P$. Observations indeed show $a^{\rm obs}_P$ larger than $a^{\rm theory}_P$ by 20\%. Thus, the theory of the present paper explains 80\% of the overall effect by the effect of the expanding geometry leaving for the thermal recoil contribution no more than 20\%.  

Second, the analysis in \citep{2011AnP...523..439R,Turyshev_2012} has been based on the usage of an entire historical record of the Pioneer tracking  telemetry. This may look as a more strong evidence of the thermal emission but the caveat is that a great care should be taken in the interpretation of the telemetry in the inner region of the solar system. Fact of the matter is that the signal of the Pioneer anomalous acceleration is strongly contaminated by the noise caused by the solar wind and radiation before the spacecraft reached the orbit of Jupiter. To avoid dealing with this treachery region of the solar system, Anderson et al \cite{2002PhRvD..65h2004A} excluded the contaminated part of the spacecraft telemetry from their analysis. Extracting a signal ($S$) from noise ($N$) in the case when the signal-to-noise ratio $S/N\ll 1$, is a severely ambiguous computational problem which can result in systematic bias in the numerical value of the fitting parameters including $a^{\rm obs}_P$. For this reason, the paper by Anderson et al \cite{2002PhRvD..65h2004A} continues to remain one of the most credible source of unbiased information about the magnitude and direction of various forces contributing to the Pioneer anomaly. 

The papers \citep{2011AnP...523..439R,Turyshev_2012} provide a firm support to the reality of the Pioneer effect and a reasonable justification that the thermal emission is a part it because of the tiny, marginally measurable secular decrease of the magnitude of the effect. However, this justification cannot be considered as a proof that the thermal recoil explains 100\% of the overall effect which is unrealistic conclusion that is not supported by the geometric theory of the Doppler tracking in cosmology presented in this paper.  The present paper along with \citep{Kopeikin_2012eph} explain the origin of the Pioneer anomaly from a simple geodesic principle of the propagation of light in the conformal geometry of the expanding FLRW universe. Our purely geometric prediction of the ``anomalous'' Doppler shift of the Pioneer spacecraft is fundamental and fully consistent with its measurement made by Anderson et al \cite{2002PhRvD..65h2004A}.

It is worth mentioning that Cassini spacecraft was also equipped with a coherent Doppler tracking system and it might be tempting to use the Cassini telemetry to measure the universal ``anomalous Cassini acceleration'' $a^{\rm theory}_C\simeq 7 \times 10^{-10}$ m$\cdot$s$^{-2}$. Unfortunately, there are large thermal and outgassing effects on Cassini that would make it difficult or impossible to say anything about the ``Cassini anomaly'' from Cassini data, during its cruise phase between Earth and Saturn \citep{2010PEPI..178..176A}. Due to the presence of the Cassini-on-board-generated systematics, the recent study \citep{2012CQGra..29w5027H} of radio science simulations in general relativity and in alternative theories of gravity is consistent with a non-detection of the ``Cassini anomalous acceleration'' effect.

\section{A resonant cavity on cosmological manifold}\lab{kk4d}

Previous section predicted the existence of the blue frequency shift of freely propagating electromagnetic wave. The question may arise what will happen with an electromagnetic wave oscillating in a microwave resonator. Will it frequency grow? If it were the case we could study the Hubble expansion of the universe in lab by measuring the drift of the microwave frequency with respect to a frequency of atomic clocks. Unfortunately, the answer is negative. This section explains in detail what exactly happens.  

Electromagnetic field in a microwave cavity resonator is governed by the homogeneous Maxwell equations in vacuum. In the main approximation the field propagates in cavity in the form of non-interacting plane waves travelling in opposite directions. A linear superposition of these waves form a set of standing waves with frequencies being determined by the cavity length. The frequencies form a discrete spectrum of electromagnetic oscillations which is calculated by solving the boundary value problem. It is possible to tune one of the cavity resonance modes to the optical band so that it can be used to interrogate the quantum transitions of an ion placed at the center of the cavity \citep{2004ESASP.554..625G}.

The frequency stability of such a clock is governed by the resonant cavity playing a role of a flywheel oscillator \citep{Tobar5168185}. Long-term stability of the clock is determined by the stability of the frequency of the quantum transitions between the different energy levels of electrons in the atoms used in the clock. The frequency instability of the cavity resonance mode is induced mainly by thermal, acoustic, mechanical, and seismic noises while the fundamental limit is due to the Brownian motion of cavity's reflecting walls \citep{2004PhRvL..93y0602N}. The cavity frequency stability is rapidly improving by making use of innovative technological designs for reducing the coupling of the cavity to the environmental disturbances \citep{2010OptCo.283.4696Z,HartLuiten}. Hence, we can expect that it will be comparable or even exceeds the stability of the atomic transitions.

Our discussion of the motion of freely-propagating photons in an expanding universe led us to the conclusion that photon's frequencies are subject to the blue shift caused by the local Hubble expansion. It is natural to extend our analysis to the case of a resonant cavity to check if this frequency drift also exists for the electromagnetic waves oscillating in the cavity. If it were true we could have a nice experimental testing of the local Hubble expansion with clocks in laboratory. We investigate this problem below and show that the result is the same as in a flat spacetime, namely, the resonant cavity {\it does not} experience any frequency shift caused by the Hubble expansion. This is because the cancellation of the Doppler shift in the superposition of two waves travelling in opposite directions in the resonant cavity.  

\subsection{Electrodynamics in FLRW Universe}\lab{meun}

Electromagnetic field tensor $F_{\a\b}=\pd_\a A_{\b}-\pd_\b A_{\a}$ is defined in terms of a vector potential $A_\a$. General-relativistic Maxwell's equations for electromagnetic vector $A^\a$ in a curved FLRW spacetime are \citep{mitowh}
\be\lab{at1}
A_\a{}^{|\b}{}_{|\b}-A^\b{}_{|\b\a}-\bar R_{\a\b}A^\b=-\frac{4\pi}{c}J_\a\;,
\ee
where $J_\a$ is a four-vector of a conserved electric current, $J^\a{}_{|\a}=0$, and $\bar R_{\a\b}=\bar g^{\m\n}\bar R_{\a\m\b\n}$ is the Ricci tensor of FLRW metric derived form (\ref{1}) and given by
\be\lab{i6a}
\bar R_{\a\b}=\frac{1}{a^2}\left[{\cal H}'\left(\bar g_{\a\b}-2\bar u_\a\bar u_\b\ri)+2\left({\cal H}^2+k\ri) \left(\bar g_{\a\b}+\bar u_\a\bar u_\b\ri)\ri]\;,
\ee
where we keep the spatial curvature $k$ arbitrary, use the conformal Hubble parameter ${\cal H}=(da/d\eta)/a$ related to the Hubble parameter $H={\cal H}/a$, and the prime denotes a time derivative with respect to the conformal time $\eta$, that is ${\cal H}'=d{\cal H}/d\eta$.

Usually, the covariant Lorentz gauge condition, $A^{\b}{}_{|\b}=0$, is imposed on vector-potential in order to eliminate the second (gauge-dependent) term in the left side of equation (\ref{at1}) and reduce it to the de Rham wave equation \citep{mitowh}. The covariant Lorentz gauge condition is not convenient in FLRW spacetime because it does not cancel the term with the Ricci tensor in (\ref{at1}). We have found \citep{Kopeikin_2012eph} that it can be eliminated if we use another gauge, $\eta^{\a\b}\pd_\a A_\b=0$ that is equivalent to
\be\lab{at2} 
A^\b{}_{|\b}=-2\H A^\b\bar u_\b\;,
\ee
or
\be\lab{at2dd} 
A^\b{}_{,\b}=+2\H A^\b\bar u_\b\;.
\ee

After making use of (\ref{at2}) and (\ref{i6a}) in (\ref{at1}) many cancellations take place, and we arrive to the exact Maxwell equations in FLRW spacetime 
\be\lab{at3} \Box A_\a=-4\pi a^2J_\a\;,
\ee
where $\Box=-\pd^2_\eta+\Delta$ is a wave operator, and $\Delta$ is the Laplace operator on the time-like hypersurface of a constant time $\eta$. In case, when the spatial curvature $k=0$, the Laplace operator is reduced to that in the Euclidean space, $\Delta=\delta^{ij}\pd_i\pd_j$. In this case $\Box$ is identical to the wave operator in Minkowski spacetime, and (\ref{at3}) is similar to Maxwell's equation for electromagnetic potential except for the scale factor, $a(t)$, in its right side. The reason for the simplicity of (\ref{at3}) is the conformal invariance of Maxwell's equations \citep{waldorf}. Solution of equation (\ref{at3}) is obtained with a standard technique of the retarded Green's function, 
\be\lab{at3a} 
A_\a(\eta,{\bm y})=\int_{\st V}\frac{a^2(s)J_a(s,{\bm y}')d^3y'}{|{\bm y}-{\bm y}'|}\;, 
\ee 
where the retarded time $s=\eta-|{\bm y}-{\bm y}'|$, and the volume integral is performed over the charge distribution bounded by the volume $V$. Equation (\ref{at3a}) tells us that the weak electromagnetic waves propagate with speed $c=1$ in the global conformal coordinates $y^\a=(\eta,{\bm y})$ of FLRW universe.

The gauge condition (\ref{at2}) applied to equation (\ref{at3a}) confirms the law of conservation of electric current, $J^\a{}_{|\a}=0$,
which is equivalent to \footnote{The law of conservation is written for the covariant components of the current that explains the sign minus in front of the second term of (\ref{at5}).}
\be\lab{at5}
\frac{\pd}{\pd \eta}\le(a^2 J_{0}\ri)-\frac{\pd}{\pd y^i}\le(a^2 J_{i}\ri)=0\;.
\ee
Hence, the total electric charge is defined by 
\be\lab{at6}
 Q=-\int_{\st V}a^2(\eta)J_0(\eta,{\bm y})d^3y\;,
\ee
and remains constant in the course of the Hubble expansion as a consequence of the conservation law (\ref{at5}). 

\subsection{Mechanical rigidity of a resonant cavity}\lab{kmz3d}

Before discussing evolution of electromagnetic field in a resonant cavity, we have to make sure that the Hubble expansion does not lead to the change of a geometric shape of the cavity. Rigidity of the cavity is determined by the chemical bonds between atoms. They are governed primarily by the Coulomb law. Hence, we have to prove that the Coulomb law preserves its classic form in the expanding FLRW universe.

For this, it is sufficient to consider only the electric potential $\phi\equiv cA_0$ of an atomic nucleus. It is given by the Taylor expansion
of the retarded argument of the vector potential (\ref{at3a}) around present time $\eta$. For the charge is conserved, it yields 
\be\lab{at3b} \phi(\eta,{\bm y})=\int_{\st V}\frac{a^2(\eta)J_0(\eta,{\bm y}')d^3y'}{|{\bm y}-{\bm y}'|}+O\le(\frac1{c^2}\ri)\;.
\ee
The integral can be expanded in terms of the electric multipoles -- the constant charge $Q$, the dipole electric moment $Q_i$, and so on, 
\be\lab{at7} 
\phi(\eta,{\bm y})=-\frac{Q}{|{\bm y}|}+\frac{Q_i y^i}{|{\bm y}|^3}+...\;.
\ee
For the sake of simplicity, we neglect the dipole and higher-order multipole moments of electric field in (\ref{at7}). Their treatment makes the calculations longer but do not change the conclusion of the present section. 

The orbital motion of an electron in a curved spacetime is governed by the second Newton's law with an electromagnetic Lorentz force in its right side taken into account \citep{mitowh}. The left side of this law is a covariant derivative from the linear momentum of electron $p^\a={m}dy^\a/d\tau$, 
\be\lab{at8} 
m\le(\frac{d^2y^a}{d\s^2}+\bar\G^\a{}_{\b\g}\frac{dy^\b}{d\s}\frac{dy^\g}{d\s}\ri)=eF^\a{}_{\b}\frac{dy^\b}{d\s}\;,
\ee
where $\s$ is the affine parameter along the electron's orbit, $e$ is the charge of electron $(e<0)$, and ${m}$ is electron's mass. The charge and mass of electron remain constant in expanding universe (see equation (\ref{at6}) above and discussion in \citep{Kopeikin_2012eph}). The Christoffel symbols in (\ref{at8}) are defined in (\ref{qq2}) and account for the Hubble expansion of FLRW universe. 
If the conformal time $\eta$ is used for parametrization of the worldline of electron, equation (\ref{at8}) assumes the following form 
\be\lab{at9} \frac{d^2y^i}{d\eta^2}+\bar\G^i{}_{\m\n}\frac{dy^\m}{d\eta}\frac{dy^\n}{d\eta}-\bar\G^0{}_{\m\n}\frac{dy^\m}{d\eta}\frac{dy^\n}{d\eta}\frac{dy^i}{d\eta}=\frac{e}{{m}}\le(F^{i}{}_\b-F^0{}_\b\frac{dy^i}{d\eta}\ri)\frac{dy^\b}{d\eta}\frac{d\s}{d\eta}\;.
\ee 
In a slow-motion approximation $d\s/d\eta=a(\eta)$, $F^i{}_0=F_{i0}/a^2$, and the electric field of the atomic nucleus is $E_i=F_{i\b}dy^\b/d\eta=\pd_i\phi$. By neglecting relativistic corrections of the order of $1/c^2$ which are negligibly small in the slow-motion approximation, we obtain the equation of orbital motion of electron in Borh's atom placed to the expanding universe 
\be\lab{at10} 
\frac{d^2{\bm y}}{d\eta^2}=-H\frac{d{\bm y}}{d\eta}+\frac{eQ}{{m}a}\frac{{\bm y}}{\rho^3}\;, 
\ee 
where $\rho=|{\bm y}|$.
Now, we employ the conformal spacetime transformation (\ref{c4}) on the worldline of the electron. In the slow-motion approximation this transformation reads
\be\lab{eqo7} \t=\int a(\eta)d\eta\;,\qquad {\bm x}=a(\eta){\bm y}\;, 
\ee 
where $\t$ is the proper time measured by the Hubble observer. Making use of (\ref{eqo7})
we can recast (\ref{at10}) to the classic form of the second Newton's law with Coulomb's force 
\be\lab{at11} {m}\frac{d^2{\bm x}}{dt^2}=\frac{eQ}{r^3}{\bm x}\;,
\ee
where $r=|{\bm x}|$.

Thus, the orbital motion of electrons in atoms, after having been expressed in physical (local) coordinates $x^\a=(\t,{\bm x})$, does not reveal any dependence on the scale factor $a(t)$ and/or the Hubble parameter $H$. This proves that the strength of the chemical bonds of the cavity's matter is not affected by the expansion of universe in the linearized Hubble approximation so that the resonant cavity is rigid and maintains its geometric shape unchanged under conditions that temperature and other environmental disturbances are kept under control. Equation (\ref{eqo7}) also tells us that the atomic frequencies are not affected by the Hubble expansion. This proves that the atomic time measured by atomic clocks, and the proper time $\t$ of the Hubble observer are identical in FLRW universe.

\subsection{The conformal metric and electromagnetic oscillations}\lab{emos5d}

Electromagnetic field in a resonant cavity obeys the homogeneous Maxwell equations that follows from (\ref{at3}),
\be\lab{eo1} 
\Box A_\a=0\;.
\ee
We assume that the walls of the cavity are ideal mirrors reflecting electromagnetic waves without dissipation so that the electromagnetic field is fully confined in the cavity. We also assume that the cavity is a rectangular box with the spatial coordinate axes of the global coordinates, $y^i=(y^1,y^2,y^3)$, directed along its sides. It allows us to solve (\ref{eo1}) by applying the method of separation of variables that splits equation (\ref{eo1}) in three, one-dimensional wave equations which are decoupled. 

The condition (\ref{at2}) bears a residual gauge freedom for the electromagnetic potential $A_\a\longrightarrow A'_\a=A_\a+\pd_\a\chi$ where $\chi$ is an arbitrary scalar function that obeys a homogeneous wave equation $\Box\chi=0$. 
The residual gauge freedom allows us to chose, $A_0=0$, in the solution of (\ref{eo1}) representing an electromagnetic wave. We further assume that the wave is propagating along $y^1\equiv y$ axis and is polarized in the $y^3$-direction so that the potential $A_i=(0,0,A)$, while the electric $E_\a=F_{\a\b}\bar u^\b$, and the magnetic $B_\a=-(1/2)\epsilon_{\a\b\m\n}F^{\m\n}\bar u^\b$, fields have the following spatial components, 
\be\lab{eo4} E_i=(0,0,E)=\le(0,0,-\frac1{a}\frac{\pd A}{\pd\eta}\ri)\;,\qquad B_i=(0,B,0)=\le(0,-\frac1{a}\frac{\pd A}{\pd x},0\ri)\;.
\ee
The field $E=E(\eta,y)$ and $B=B(\eta,y)$ satisfy the wave equation that is derived from (\ref{eo1}). In particular, the wave equation for the electric field, $E\equiv{\cal E}/a$, is 
\be\lab{eqo5}
-\frac{\pd^2 \cal E}{\pd\eta^2}+\frac{\pd^2\cal E}{\pd y^2}=0\;, 
\ee 
and a similar equation exists for the magnetic field ${\cal B}=Ba$.

The tangential components of $E_i$ must vanish on the cavity's walls that are perpendicular to $y$-axis. According to our conclusion from the previous section, the cavity keeps its geometric shape unchanged with respect to physical (local) coordinates, $x^i$, but not with respect to the conformal coordinates $y^i=x^i/a(\eta)$. Let as choose the left wall of the cavity as the origin of $x^1\equiv x$ axis so that the value of the global and local coordinates coincide at any instant of time, $x^1=y^1=0$. The right wall of the cavity is fixed with respect to the local coordinate at $x=L$ but it moves adiabatically with respect to the global coordinate of the wall, $y=l(\eta)=L/a(\eta)$. The boundary conditions imposed on the electric field at the cavity's walls are \be\lab{nbz3d}{\cal E}[\eta,y=0]=0\qquad,\qquad {\cal E}[\eta,y=l(\eta)]=0\;.\ee
Standing wave solution of the second-order partial differential equation (\ref{eqo5}) with the adiabatic time-dependent boundary condition imposed on the electric field, has been worked out in a number of papers, for example, in \citep{1970JMP....11.2679M,1994PhRvL..73.1931L,1996PhRvA..53.2664D,1998PhRvA..57.4784J}. It consists of a discrete spectrum of standing waves where each harmonic can be represented as a Fourier series with respect to the {\it instantaneous} basis
\be\lab{eq6}
{\cal E}(\eta,y)=\sum_{n=1}^\infty Q_n(\eta)\sin\le[\frac{\pi n y}{l(\eta)}\ri]\;.
\ee 
This form of the solution satisfies the boundary conditions exactly. 

We put (\ref{eq6}) to the wave equation (\ref{eqo5}), multiply it with $\sin\le[\pi m y/l(t)\ri]$ ($n\not= m)$, and integrate over $y$ from $y=0$ to $y=l(\eta)$. It yields the ordinary differential equation for the amplitude $Q_n=Q_n(\eta)$ of the modes of the standing wave that reads  
\be\lab{eqo12} Q''_n+\omega_n^2(\eta){Q}_n=+2{\cal H}\sum_{k=1}^\infty w_{nk}\omega_kQ'_k\;, \ee 
where the prime denotes the time derivative with respect to the conformal time $\eta$, the (time-dependent) resonant frequencies 
\be\lab{eq12a} \omega_n(\eta)=a(\eta)\Omega_n\;, \ee 
$\Omega_n\equiv \pi n/L$ is a constant frequency, and 
\be\lab{eqo13} w_{nk}=\frac2{L}\int^L_0\xi\sin\le(\Omega_n \xi\ri)\cos\le(\Omega_k \xi\ri)d\xi\;, \ee 
is a set of constant coefficients.

The scale factor $a(\eta)$ changes very slowly due to the Hubble expansion while the frequency of the electromagnetic oscillations in the cavity is very high. For this reason, equation (\ref{eqo12}) can be solved in an adiabatic approximation that neglects all terms with the time derivatives of $Q_k$ in its right side \citep{wkb_approx}. The adiabatic solution of (\ref{eqo12}) for the standing wave is
\be\lab{eq14}
Q_n(\eta)=\frac{b_n}{\sqrt{a(\eta)}}\cos\le[\int\limits_0^\eta\omega_n(\eta)d\eta+\phi_n\ri]\;,
\ee
where $b_n$ and $\phi_n$ are constant amplitude and phase of the $n$-th mode of the harmonic oscillations.
We notice that in the adiabatic approximation, the product $\int_0^\eta\omega_n(\eta)d\eta=\Omega_n \t$, where $\t=t$ is the proper time of the Hubble observer. Therefore, the electric field in the cavity is
\be\lab{bv3x1}
E(\t,x)=\sum_{n=1}^\infty B_n(\t)\cos\left(\Omega_n\t+\phi_n\ri)\sin\le(\frac{\pi n x}{L}\ri)\;,
\ee
where we have used transformation from the global to local spatial coordinates $x=ya(\eta)$ and introduced the notation $B_n(\t)= b_na^{-3/2}$ for slowly-changing amplitude of the oscillations.

Equation (\ref{bv3x1}) demonstrates that the frequency of electromagnetic oscillations in a resonant cavity is not subject to the Hubble expansion because the frequencies, $\Omega_n$, are constant with respect to the proper time $\t$ of the Hubble observer measured with the help of an atomic clock. This theoretical conclusion has got a rather precise experimental verification \citep{Storz_1998OptL}. The amplitude $A_n(\t)$ of the standing wave in (\ref{bv3x1}) depends on time due to the Hubble expansion through the time-dependent scale factor $a=a(\t)$. This change in the amplitude of the electric field is minuscule on any (practically reasonable) observational time span and can be safely neglected.

\section{Final comments}\lab{opwd}
\begin{enumerate}
\item We have build the LIC by applying the special conformal transformation (\ref{c4}). Comparison with other approaches \citep{hongya:1920,hongya:1924,2005ESASP.576..305K,2007CQGra..24.5031M,2010RvMP...82..169C} to build the LIC in cosmology reveals that all of them bring about the same coordinate transformation (\ref{c4}) in the linearized Hubble approximation. Therefore, there is no difference which approach is used to build the local coordinates if the quadratic terms in the Hubble parameter are not considered. Our approach to build LIC helps to realize that the transformation to the local coordinates on the expanding cosmological manifold is, in fact, an infinitesimal special conformal transformation. 

\item  Introducing a local physical distance $x^i=R(t)y^i$, recast (\ref{1}) to the following form
\be\lab{mzd4}
ds^2=-\le(1+H^2{\bm x}^2\ri)dt^2-2Hx^idx^idt+\d_{ij}dx^idx^j\;,
\ee
which can be written down as
\be\lab{mb3df}
ds^2=-dt^2+\d_{ij}\left(dx^i-X^idt\ri)\left(dx^j-X^jdt\ri)\;,
\ee
where vector field $X^i\equiv Hx^i$. Metric (\ref{mb3df}) is exactly the warp-drive metric that was suggested by Alcubierre \citep{warp_drive} to circumvent the light-speed limit in general relativity. All mathematical properties of the warp-drive metric that have been analysed, for example, in \citep{warp_drive2002} are valid in the local coordinates $(t,x^i)$ where $t$ is the proper time $\t$ of the local static observers ($x^i={\rm const.}$), if we neglect terms of the order of $H^2$. The metric (\ref{mzd4}) is non-inertial but it can be converted to the flat Minkowski metric in a neighbourhood of the coordinate origin with the help of an additional transformation of the proper time $t=\t$ to a local time coordinate $x^0$ as shown in (\ref{c4}). The local time coordinate $x^0$ coincides with the proper time $\t$ of the static observers but deviates by a scale factor $a(\t)$ from $\t$ on the light cone. This is why the optical metric (\ref{fe3}) emerges in the local reference frame.

\item The origin of the optical metric (\ref{fe3}) in the local frame can be tracked back to FLRW metric (\ref{1}) where we take $k=0$ for simplicity. We, first, convert it to the form (\ref{mzd4}) and, then, project it onto the null cone hypersurface which is defined in the leading order approximation by equation ${\bm x}^2-t^2=0$. Taking a differential from this equation yields, $x^idx^i-tdt=0$. We make use this differential relation along with ${\bm x}^2=t^2$ in the warp-drive metric (\ref{mzd4}) which project it to the metric on the null cone hypersurface
\be\lab{m4zs}
ds^2=-\le(1+Ht\ri)^2dt^2+\d_{ij}dx^idx^j\;.
\ee
This metric exactly coincides with the optical metric given in (\ref{fe3}) after introducing the scale factor $a(t)=1+Ht$ and  identification of the cosmological time $t$ with the proper time $\t$ of the static observers having $x^i={\rm const}$. We remind the reader that the optical metric (\ref{m4zs}) is valid only on the light cone that is a degenerated hypersurface in the sense that all equations derived from (\ref{m4zs}) are valid under condition of $ds^2=0$.

\item The reader can wonder if there arises birefringence of the wave vector in the optical metric because the speed of light propagating along radial geodesics, depends on where it moves to or out of the coordinate origin as have been discussed in Section \ref{d4d}. To answer this question, we notice that in the local frame of the Hubble observer the optical metric (\ref{fe3}) yields the index of refraction, $n=1/a(\tau)=1-H\t$. It does not depend on spatial coordinates. Hence, a light ray moving in a particular direction will not experience birefringence. Nevertheless, if one makes a Lorentz transformation to another frame, moving with a constant velocity ${\bf v}={\rm v}^i$ with respect to the Hubble flow, the index of refraction becomes (we remind that $c=1$)
\be\lab{inrf1}
n=1-H\frac{\t-{\bf v}\cdot{\bm x}}{1-{\bf v}^2}\simeq 1-H\t+H{\bf v}\cdot{\bm x}\;.
\ee
It depends on the spatial coordinates of the light ray and, hence, birefringence will be formally present. This is because the FLRW metric is homogeneous and isotropic only with respect to the frame formed by the congruence of the Hubble observers. In any other frame moving with a constant velocity ${\rm v}^i$ with respect to the Hubble flow, there is a preferred direction in space given by the velocity vector ${\rm v}^i$. This direction determines the optical axis. The maximal difference between the two values of the refractive index for the light rays propagating in the direction of the optical axis and perpendicular to it, is $\delta=({\rm v}/c)H\t$, where $c$ is the speed of light, and it changes as light propagates. The birefringence effect caused by motion of the Solar system with respect to the Hubble flow is extremely small because ${\rm v}/c\ll 1$. There are other reasons to expect birefringes in cosmology, for example, cause by the possible violation of the EEP due to a pseudo-scalar fields \citep{Ni_2008PThPS}. This violation would lead to the rotation of the plane of linear polarization of radiation traveling over cosmological distances, the so-called Cosmological Birefringence (CB) \citep{sperello_2011}. This subject has been recently revived by the possibility of testing for CB using the CMB polarization.

\item The analysis of EEP given in the present paper, was focused on the solar system experiments as contrasted with pure cosmological tests. There are other possible tests which can be potentially conducted for testing the formalism worked out in the present paper, for example, with binary pulsars \citep{krawex_2009}. Timing measurements establish a very precise local frame for the binary pulsar system which is not affected by the Hubble expansion as explained in \citep{Kopeikin_2012eph}. On the other hand, we expect that the cosmological expansion influence the time of propagation of radio pulses from the pulsar to observer on Earth, and this effect should be seen in the secular change of the orbital period $P_b$ of binary pulsars of the order of $\dot P_b/P_b=H\simeq 2.3\times 10^{-18}$. This effect is superimposed on the effect of the orbital decay due to the emission of gravitational waves by the binary system and introduces a bias to the observed value of $\dot P_b$ in addition to the Shklovskii effect \citep{Damour_1991ApJ}. However, the orbital decay of binary pulsars with wide orbits is negligible small, hence, allowing to observer the Hubble expansion effect in the secular change of the orbital period. 
\end{enumerate}

\appendix
\section{Comparison with the paper ``Influence of global cosmological expansion on local dynamics and kinematics'' by M. Carrera and D. Giulini}

The present consensus seems to be that the Hubble exapnsion cannot be observed in any kind of local experiments. The paper by M. Carrera and D. Giulini \citep{2010RvMP...82..169C} summarizes this point of view. We believe the consensus is premature and based on the insufficiently deep understanding of the problem of construction of the local reference frames in cosmology and the principles of astronomical measurements in the solar system \citep{kopeikin_2011book}. This appendix comments on the paper \citep{2010RvMP...82..169C} with regard to the Doppler effect in cosmology as measured by the local observers. 

M. Carrera and D. Giulini \citep{2010RvMP...82..169C} (see also \citep{2006CQGra..23.7483C}) considered the Doppler effect in the local frame of FLRW universe. They claim that the cosmological expansion does not produce any noticeable Doppler effect in the local radio-wave frequency measurements. Unfortunately, they misidentified the canonical parameter $\lambda$ on the light cone characteristic with the proper time $\t$ measured by the static observers. According to \citep{2010RvMP...82..169C} the metric on the light cone written in the local Gaussian coordinates coincides with the Minkowski metric at any instant of time $\t$ measured along the light-ray path by the static observers. This is impossible due to the specific mathematical structure of the local conformal diffeomorphism as we have explained in section (\ref{opm123}).

Moreover, Carrera and Giulini \citep{2010RvMP...82..169C} did their calculation of the Doppler shift not in the reference frame based on the local static observers but in the reference frame based on congruence of the Hubble observers. We notice that there is no any practical way to materialize the local frame with the Hubble observers which remain an abstract mathematical construction in the solar system as contrasted to the static observers materializing the normal Gaussian coordinates in the local frame which we built in section \ref{sec4}. Nonetheless, it is possible to derive the Doppler shift for the Hubble observers from a pure mathematical point of view. It is clear that the result for the Doppler shift measured by these observers, will be a standard (red shift) formula employed in cosmology for distant quasars \citep{waldorf}
\be\lab{d5cg}
\frac{\omega_2}{\omega_1}=\frac{a(\t_1)}{a(\t_2)}=1-H(\t_2-\t_1)\;.
\ee 
The last term in right side of (\ref{d5cg}) differs from our (blue shift) formula (\ref{d5a}) for the Doppler shift in the local frame by sign because it refers to the different set of observers. Making use of tetrad formalism, Carerra and Giulini extrapolated (\ref{d5cg}) to the local frame of the static observers, which move with respect to the Hubble observers with the (negative) velocity of the Hubble flow. It introduces to (\ref{d5cg}) an additional term -- a kinematic Doppler shift of a special relativistic origin -- which cancels the Hubble red shift term in the right side of (\ref{d5cg}). It made Carerra and Giulini convinced that the Hubble expansion does not affect the frequency of light in the local frame. 

This part of the calculation in \cite{2010RvMP...82..169C} is formally correct but it does not distinguish between the value of the local time coordinate $x^0=\t$ taken at the point of emission of light and the value of this coordinate at the point of reception which is lying on the light cone, and hence, for the light reception point $x^0=\int a(\t)d\t=\t+(H/2)\t^2$ as we showed in section \ref{opm123}. Carrera and Giulini interpreted the local time coordinate $x^0$ on the light cone as equal to the proper time $\t$ for both the emitter and the receiver of light but this is not a valid procedure as the coordinate time $x^0$ coincides with the proper time $\t$ only for static observers but behaves differently as a function of $\t$ for moving particles. The appropriate relation between the local time coordinate $x^0$ and the proper time $\t$ on the light cone is what Carerra and Giulini have been missing. Correct parametrization of the light ray trajectory with the parameter $x^0=\t+(H/2)\t^2$ made in addition to the tetrad transformation, converts (\ref{d5cg}) to our equation (\ref{d5a}) predicting the existence of the blue frequency shift of light measured in cosmology by the static observers. 

Because Carrera and Giulini used the reference frame based on the Hubble observers, they found out that the radar distance $\ell$ between the Hubble observers is affected by the Hubble expansion and changes as time goes on \citep[Eq. 23]{2006CQGra..23.7483C}
\be\lab{ccr8b}
\frac{d\ell}{d\t}=1+H\t;.
\ee 
This is again a standard result of cosmological theory \citep{waldorf} which is inapplicable to the local astronomical measurements conducted by reference observers being at rest in the local frame. As we showed in section \ref{d4d}, the radar distance $\ell=r-Hr^2$ between the static observers does not change as the universe expands because the coordinate distance $r$ between the static observers does not change. 

\newpage
\acknowledgments
I am grateful to S. di Serego Alighieri, J.~D. Anderson, Yu. V. Baryshev, M. Yu. Khlopov, S.~A. Klioner, S.~A. Levshakov, B. Mashhoon, F. Melia, V. F. Mukhanov, B. Schutz, S. Satpathy, M. Soffel, O. Titov, S. Schiller and M. Tobar for valuable discussions and critical comments. 

\newpage

\bibliographystyle{ieeetr}
\bibliography{kopeikin_bib}

\begin{thebibliography}{10}

\bibitem{baryshev_2012ASSL}
Y.~{Baryshev} and P.~{Teerikorpi}, {\em {Fundamental Questions of Practical
  Cosmology}}, vol.~383 of {\em Astrophysics and Space Science Library}.
\newblock Springer, Berlin, 2012.

\bibitem{2006LRR.....9....3W}
C.~M. {Will}, ``{The Confrontation between General Relativity and
  Experiment},'' {\em Living Reviews in Relativity}, vol.~9, p.~3, Mar. 2006.

\bibitem{kopeikin_2011book}
S.~{Kopeikin}, M.~{Efroimsky}, and G.~{Kaplan}, {\em {Relativistic Celestial
  Mechanics of the Solar System}}.
\newblock Weinheim: Wiley-VCH, Sept. 2011.

\bibitem{einstein1961relativity}
A.~Einstein, {\em Relativity, the special and the general theory: a popular
  exposition by Albert Einstein}.
\newblock Crown Publishers, 1961.

\bibitem{Einstein_P1916}
A.~Einstein, ``{$\rm\ddot{U}$ber Friedrich Kottlers Abhandlung
  ``$\rm\ddot{U}$ber Einsteins $\rm\ddot{A}$quivalenzhypothese und die
  Gravitation''},'' {\em Annalen der Physik}, vol.~356, no.~22, pp.~639--642,
  1916.

\bibitem{Norton_1993}
J.~D. {Norton}, ``{General covariance and the foundations of general
  relativity: eight decades of dispute},'' {\em Reports on Progress in
  Physics}, vol.~56, pp.~791--858, July 1993.

\bibitem{will_1993}
C.~M. {Will}, {\em {Theory and Experiment in Gravitational Physics}}.
\newblock Cambridge: Cambridge University Press, 396 pp., Mar. 1993.

\bibitem{2005pfc..book.....M}
V.~{Mukhanov}, {\em {Physical Foundations of Cosmology}}.
\newblock Cambridge: Cambridge University Press, Nov. 2005.

\bibitem{weinberg_2008}
S.~Weinberg, {\em {Cosmology}}.
\newblock Oxford University Press, 2008.

\bibitem{wmap_2011}
N.~Jarosik, C.~L. Bennett, J.~Dunkley, B.~Gold, M.~R. Greason, M.~Halpern,
  R.~S. Hill, G.~Hinshaw, A.~Kogut, E.~Komatsu, D.~Larson, M.~Limon, S.~S.
  Meyer, M.~R. Nolta, N.~Odegard, L.~Page, K.~M. Smith, D.~N. Spergel, G.~S.
  Tucker, J.~L. Weiland, E.~Wollack, and E.~L. Wright, ``Seven-year wilkinson
  microwave anisotropy probe (wmap) observations: Sky maps, systematic errors,
  and basic results,'' {\em The Astrophysical Journal Supplement Series},
  vol.~192, no.~2, p.~14, 2011.

\bibitem{2007JMP....48l2501I}
M.~{Ibison}, ``{On the conformal forms of the Robertson-Walker metric},'' {\em
  Journal of Mathematical Physics}, vol.~48, p.~122501, Dec. 2007.

\bibitem{waldorf}
R.~M. {Wald}, {\em {General relativity}}.
\newblock Chicago: University of Chicago Press, 1984.

\bibitem{schmidt_1987}
H.~Friedrich and B.~G. Schmid, ``Conformal geodesics in general relativity,''
  {\em Proceedings of the Royal Society of London. Series A, Mathematical and
  Physical Sciences}, vol.~414, no.~1846, pp.~pp. 171--195, 1987.

\bibitem{2010RvMP...82..169C}
M.~{Carrera} and D.~{Giulini}, ``{Influence of global cosmological expansion on
  local dynamics and kinematics},'' {\em Reviews of Modern Physics}, vol.~82,
  pp.~169--208, Jan. 2010.

\bibitem{Kopeikin_2012eph}
S.~M. {Kopeikin}, ``{Celestial ephemerides in an expanding universe},'' {\em
  \prd}, vol.~86, p.~064004, Sept. 2012.

\bibitem{2013strs.book.....S}
M.~{Soffel} and R.~{Langhans}, {\em {Space-Time Reference Systems}}.
\newblock 2013.

\bibitem{Rohrlich1963169}
F.~{Rohrlich}, ``The principle of equivalence,'' {\em Annals of Physics},
  vol.~22, no.~2, pp.~169 -- 191, 1963.

\bibitem{Harvey1964383}
A.~Harvey, ``The principle of equivalence,'' {\em Annals of Physics}, vol.~29,
  no.~3, pp.~383 -- 390, 1964.

\bibitem{kopeikin_2013}
S.~{Kopeikin}, ``{Equivalence Principle in Cosmology},'' {\em ArXiv e-prints},
  July 2013.

\bibitem{hongya:1920}
L.~Hongya, ``{Cosmological models in globally geodesic coordinates. I.
  Metric},'' {\em Journal of Mathematical Physics}, vol.~28, no.~8,
  pp.~1920--1923, 1987.

\bibitem{hongya:1924}
L.~Hongya, ``{Cosmological models in globally geodesic coordinates. II.
  Near-field approximation.},'' {\em Journal of Mathematical Physics}, vol.~28,
  no.~8, pp.~1924--1927, 1987.

\bibitem{2005ESASP.576..305K}
S.~A. {Klioner} and M.~H. {Soffel}, ``{Refining the Relativistic Model for
  Gaia: Cosmological Effects in the BCRS},'' in {\em The Three-Dimensional
  Universe with Gaia} (C.~{Turon}, K.~S. {O'Flaherty}, and M.~A.~C. {Perryman},
  eds.), vol.~576 of {\em ESA Special Publication}, pp.~305--308, Jan. 2005.

\bibitem{2007CQGra..24.5031M}
B.~{Mashhoon}, N.~{Mobed}, and D.~{Singh}, ``{Tidal dynamics in cosmological
  spacetimes},'' {\em Classical and Quantum Gravity}, vol.~24, pp.~5031--5046,
  Oct. 2007.

\bibitem{nizim}
W.-T. {Ni} and M.~{Zimmermann}, ``{Inertial and gravitational effects in the
  proper reference frame of an accelerated, rotating observer},'' {\em \prd},
  vol.~17, pp.~1473--1476, Mar. 1978.

\bibitem{Mielke_1977}
E.~W. {Mielke}, ``{Conformal changes of metrics and the initial-value problem
  of general relativity},'' {\em General Relativity and Gravitation}, vol.~8,
  pp.~321--345, May 1977.

\bibitem{BMS_group}
T.~M. Adamo, E.~T. Newman, and C.~Kozameh, ``Null geodesic congruences,
  asymptotically-flat spacetimes and their physical interpretation,'' {\em
  Living Reviews in Relativity}, vol.~15, no.~1, 2012.

\bibitem{1966PhRv..150.1183K}
H.~A. {Kastrup}, ``{Gauge Properties of the Minkowski Space},'' {\em Physical
  Review}, vol.~150, pp.~1183--1193, Oct. 1966.

\bibitem{cft_intro}
M.~{Schottenloher}, {\em {A Mathematical Introduction to Conformal Field
  Theory}}, vol.~759 of {\em Lecture Notes in Physics}.
\newblock Springer Verlag: Berlin, 2008.

\bibitem{1962AnP...464..388K}
H.~A. {Kastrup}, ``{Zur physikalischen Deutung und darstellungstheoretischen
  Analyse der konformen Transformationen von Raum und Zeit},'' {\em Annalen der
  Physik}, vol.~464, pp.~388--428, 1962.

\bibitem{eps_1972}
J.~{Ehlers}, F.~A.~E. {Pirani}, and A.~{Schild}, ``{Republication of: The
  geometry of free fall and light propagation},'' {\em General Relativity and
  Gravitation}, vol.~44, pp.~1587--1609, June 2012.

\bibitem{Gordon_1923}
W.~{Gordon}, ``{Zur Lichtfortpflanzung nach der Relativit{\"a}tstheorie},''
  {\em Annalen der Physik}, vol.~377, pp.~421--456, 1923.

\bibitem{Synge_GRbook}
J.~L. {Synge}, {\em {Relativity: The General Theory}}.
\newblock Series in Physics, Amsterdam: North-Holland Publication Co., 1964.

\bibitem{Ehlers_1967}
J.~{Ehlers}, ``{Zum {\"U}bergang von der Wellenoptikzur geometrischen Optik in
  der allgemeinen Relativit{\"a}tstheorie},'' {\em Zeitschrift Naturforschung
  Teil A}, vol.~22, p.~1328, Sept. 1967.

\bibitem{perlick_2006}
V.~{Perlick}, ``{Fermat principle in Finsler spacetimes},'' {\em General
  Relativity and Gravitation}, vol.~38, pp.~365--380, Feb. 2006.

\bibitem{Perlick_2000}
V.~{Perlick}, {\em {Ray Optics, Fermat's Principle, and Applications to General
  Relativity}}.
\newblock Springer-Verlag, 2000.

\bibitem{2012PhRvD..86l4024N}
M.~{Novello} and E.~{Bittencourt}, ``{Gordon metric revisited},'' {\em \prd},
  vol.~86, p.~124024, Dec. 2012.

\bibitem{PhysRevD.83.084047}
S.~Casey, M.~Dunajski, G.~Gibbons, and C.~Warnick, ``Optical metrics and
  projective equivalence,'' {\em Phys. Rev. D}, vol.~83, p.~084047, Apr 2011.

\bibitem{Kantowski_2009}
B.~Chen and R.~Kantowski, ``Including absorption in gordon's optical metric,''
  {\em Phys. Rev. D}, vol.~79, p.~104007, May 2009.

\bibitem{lrr-2005-12}
C.~Barceló, S.~Liberati, and M.~Visser, ``Analogue gravity,'' {\em Living
  Reviews in Relativity}, vol.~8, no.~12, 2005.

\bibitem{1992Natur.356..207F}
E.~{Fischbach} and C.~{Talmadge}, ``{Six years of the fifth force},'' {\em
  \nat}, vol.~356, pp.~207--215, Mar. 1992.

\bibitem{1999CQGra..16.1313B}
W.~B. {Bonnor}, ``{Size of a hydrogen atom in the expanding universe},'' {\em
  Classical and Quantum Gravity}, vol.~16, pp.~1313--1321, Apr. 1999.

\bibitem{2006PhRvD..74f4019C}
C.~{Chicone} and B.~{Mashhoon}, ``{Explicit Fermi coordinates and tidal
  dynamics in de Sitter and G{\"o}del spacetimes},'' {\em \prd}, vol.~74,
  p.~064019, Sept. 2006.

\bibitem{2009homp.book..365S}
R.~{Sch{\"o}del}, ``{Length and Size},'' in {\em Handbook of Optical Metrology:
  Principles and Applications} (T.~{Yoshizawa}, ed.), pp.~365--392, CRC Press,
  Taylor \& Francis Group, Boca Raton, FL USA, Feb. 2009.

\bibitem{iau:2012}
G.~H. {Kaplan}, C.~Y. {Hohenkerk}, T.~{Fukushima}, J.-E. {Arlot}, J.~A.
  {Bangert}, S.~A. {Bell}, W.~M. {Folkner}, M.~{Lara}, E.~V. {Pitjeva}, S.~E.
  {Urban}, and J.~{Vondr{\'a}k}, ``{Commission 4: Ephemerides},'' {\em
  Transactions of the International Astronomical Union, Series A}, vol.~28,
  pp.~9--14, Apr. 2012.

\bibitem{Williams_2012}
J.~G. {Williams}, S.~G. {Turyshev}, and D.~H. {Boggs}, ``{Lunar laser ranging
  tests of the equivalence principle},'' {\em Classical and Quantum Gravity},
  vol.~29, p.~184004, Sept. 2012.

\bibitem{LanLif}
L.~D. Landau and E.~M. Lifshitz, {\em {The classical theory of fields}}.
\newblock Oxford: Pergamon Press, 1975.

\bibitem{Folkner_2009IPNPR}
W.~M. {Folkner}, J.~G. {Williams}, and D.~H. {Boggs}, ``{The Planetary and
  Lunar Ephemeris DE 421},'' {\em Interplanetary Network Progress Report},
  vol.~178, p.~C1, Aug. 2009.

\bibitem{Fienga_2009AA}
A.~{Fienga}, J.~{Laskar}, T.~{Morley}, H.~{Manche}, P.~{Kuchynka}, C.~{Le
  Poncin-Lafitte}, F.~{Budnik}, M.~{Gastineau}, and L.~{Somenzi}, ``{INPOP08, a
  4-D planetary ephemeris: from asteroid and time-scale computations to ESA
  Mars Express and Venus Express contributions},'' {\em \aap}, vol.~507,
  pp.~1675--1686, Dec. 2009.

\bibitem{murthyetal08}
T.~W. {Murphy}, E.~G. {Adelberger}, J.~B.~R. {Battat}, L.~N. {Carey}, C.~D.
  {Hoyle}, P.~{Leblanc}, E.~L. {Michelsen}, K.~{Nordtvedt}, A.~E. {Orin}, J.~D.
  {Strasburg}, C.~W. {Stubbs}, H.~E. {Swanson}, and E.~{Williams}, ``{The
  Apache Point Observatory Lunar Laser-ranging Operation: Instrument
  Description and First Detections},'' {\em \pasp}, vol.~120, pp.~20--37, Jan.
  2008.

\bibitem{weinberg_1972}
S.~{Weinberg}, {\em {Gravitation and Cosmology: Principles and Applications of
  the General Theory of Relativity}}.
\newblock New York: John Wiley \& Sons, Inc., July 1972.

\bibitem{mukh_book}
V.~{Mukhanov}, {\em {Physical Foundations of Cosmology}}.
\newblock Cambridge: Cambridge University Press, 442 pp., Nov. 2005.

\bibitem{2012IAUJD...7E..40C}
N.~{Capitaine}, S.~{Klioner}, and D.~{McCarthy}, ``{The re-definition of the
  astronomical unit of length: reasons and consequences},'' in {\em IAU Joint
  Discussion}, IAU Joint Discussion, 2012.

\bibitem{2002PhRvD..65f4025K}
S.~{Kopeikin} and B.~{Mashhoon}, ``{Gravitomagnetic effects in the propagation
  of electromagnetic waves in variable gravitational fields of arbitrary-moving
  and spinning bodies},'' {\em \prd}, vol.~65, p.~064025, Mar. 2002.

\bibitem{2006CQGra..23.7483C}
M.~{Carrera} and D.~{Giulini}, ``{On Doppler tracking in cosmological
  spacetimes},'' {\em Classical and Quantum Gravity}, vol.~23, pp.~7483--7492,
  Dec. 2006.

\bibitem{2002PhRvD..65h2004A}
J.~D. {Anderson}, P.~A. {Laing}, E.~L. {Lau}, A.~S. {Liu}, M.~M. {Nieto}, and
  S.~G. {Turyshev}, ``{Study of the anomalous acceleration of Pioneer 10 and
  11},'' {\em \prd}, vol.~65, p.~082004, Apr. 2002.

\bibitem{2007AdSpR..39..291T}
S.~G. {Turyshev}, M.~M. {Nieto}, and J.~D. {Anderson}, ``{Lessons learned from
  the Pioneers 10/11 for a mission to test the Pioneer anomaly},'' {\em
  Advances in Space Research}, vol.~39, pp.~291--296, 2007.

\bibitem{2011AnP...523..439R}
B.~{Rievers} and C.~{L{\"a}mmerzahl}, ``{High precision thermal modeling of
  complex systems with application to the flyby and Pioneer anomaly},'' {\em
  Annalen der Physik}, vol.~523, pp.~439--449, June 2011.

\bibitem{Turyshev_2012}
S.~G. {Turyshev}, V.~T. {Toth}, G.~{Kinsella}, S.-C. {Lee}, S.~M. {Lok}, and
  J.~{Ellis}, ``{Support for the Thermal Origin of the Pioneer Anomaly},'' {\em
  Physical Review Letters}, vol.~108, p.~241101, June 2012.

\bibitem{2010PEPI..178..176A}
J.~D. {Anderson} and G.~{Schubert}, ``{Rhea's gravitational field and interior
  structure inferred from archival data files of the 2005 Cassini flyby},''
  {\em Physics of the Earth and Planetary Interiors}, vol.~178, pp.~176--182,
  Feb. 2010.

\bibitem{2012CQGra..29w5027H}
A.~{Hees}, B.~{Lamine}, S.~{Reynaud}, M.-T. {Jaekel}, C.~{Le Poncin-Lafitte},
  V.~{Lainey}, A.~{F{\"u}zfa}, J.-M. {Courty}, V.~{Dehant}, and P.~{Wolf},
  ``{Radioscience simulations in general relativity and in alternative theories
  of gravity},'' {\em Classical and Quantum Gravity}, vol.~29, p.~235027, Dec.
  2012.

\bibitem{2004ESASP.554..625G}
P.~{Gill}, G.~P. {Barwood}, H.~A. {Klein}, G.~{Huang}, H.~S. {Margolis}, S.~N.
  {Lea}, M.~{Oxborrow}, and S.~A. {Webster}, ``{Optical clocks and ultra-stable
  optical oscillators for navigation, space science and astronomy},'' in {\em
  5th International Conference on Space Optics} (B.~{Warmbein}, ed.), vol.~554
  of {\em ESA Special Publication}, pp.~625--630, June 2004.

\bibitem{Tobar5168185}
J.~Millo, Y.~Le~Coq, S.~Bize, J.~Guena, H.~Jiang, M.~Abgrall, E.~English,
  A.~Clairon, G.~Santarelli, and M.~Tobar, ``Flywheel oscillator for atomic
  fountain clocks using ultra-stable lasers and a fiber-based optical frequency
  comb,'' in {\em Frequency Control Symposium, 2009 Joint with the 22nd
  European Frequency and Time forum. IEEE International}, pp.~280 --281, april
  2009.

\bibitem{2004PhRvL..93y0602N}
K.~{Numata}, A.~{Kemery}, and J.~{Camp}, ``{Thermal-Noise Limit in the
  Frequency Stabilization of Lasers with Rigid Cavities},'' {\em Physical
  Review Letters}, vol.~93, p.~250602, Dec. 2004.

\bibitem{2010OptCo.283.4696Z}
Y.~N. {Zhao}, J.~{Zhang}, J.~{Stuhler}, G.~{Schuricht}, F.~{Lison}, Z.~H. {Lu},
  and L.~J. {Wang}, ``{Sub-Hertz frequency stabilization of a commercial diode
  laser},'' {\em Optics Communications}, vol.~283, pp.~4696--4700, Dec. 2010.

\bibitem{HartLuiten}
J.~G. Hartnett and A.~N. Luiten, ``\textit{Colloquium}: Comparison of
  astrophysical and terrestrial frequency standards,'' {\em Rev. Mod. Phys.},
  vol.~83, pp.~1--9, Jan 2011.

\bibitem{mitowh}
C.~W. {Misner}, K.~S. {Thorne}, and J.~A. {Wheeler}, {\em {Gravitation}}.
\newblock San Francisco: W.H.~Freeman and Co., 1973.

\bibitem{1970JMP....11.2679M}
G.~T. {Moore}, ``{Quantum Theory of the Electromagnetic Field in a
  Variable-Length One-Dimensional Cavity},'' {\em Journal of Mathematical
  Physics}, vol.~11, pp.~2679--2691, Sept. 1970.

\bibitem{1994PhRvL..73.1931L}
C.~K. {Law}, ``{Resonance response of the quantum vacuum to an oscillating
  boundary},'' {\em Physical Review Letters}, vol.~73, pp.~1931--1934, Oct.
  1994.

\bibitem{1996PhRvA..53.2664D}
V.~V. {Dodonov} and A.~B. {Klimov}, ``{Generation and detection of photons in a
  cavity with a resonantly oscillating boundary},'' {\em \pra}, vol.~53,
  pp.~2664--2682, Apr. 1996.

\bibitem{1998PhRvA..57.4784J}
M.~{Janowicz}, ``{Evolution of wave fields and atom-field interactions in a
  cavity with one oscillating mirror},'' {\em \pra}, vol.~57, pp.~4784--4790,
  June 1998.

\bibitem{wkb_approx}
R.~Bellman and R.~Vasudevan, ``Eikonal equation and the wkb approximation,'' in
  {\em Wave Propagation}, vol.~17 of {\em Mathematics and Its Applications},
  pp.~21--37, Springer Netherlands, 1985.

\bibitem{Storz_1998OptL}
R.~{Storz}, C.~{Braxmaier}, K.~{J{\"a}ck}, O.~{Pradl}, and S.~{Schiller},
  ``{Ultrahigh long-termdimensional stability of a sapphire cryogenic optical
  resonator},'' {\em Optics Letters}, vol.~23, pp.~1031--1033, July 1998.

\bibitem{warp_drive}
M.~{Alcubierre}, ``{LETTER TO THE EDITOR: The warp drive: hyper-fast travel
  within general relativity},'' {\em Classical and Quantum Gravity}, vol.~11,
  pp.~L73--L77, May 1994.

\bibitem{warp_drive2002}
J.~{Nat{\'a}rio}, ``{Warp drive with zero expansion},'' {\em Classical and
  Quantum Gravity}, vol.~19, pp.~1157--1165, Mar. 2002.

\bibitem{Ni_2008PThPS}
W.-T. {Ni}, ``{From Equivalence Principles to Cosmology: Cosmic Polarization
  Rotation, CMB Observation, Neutrino Number Asymmetry, Lorentz Invariance and
  CPT},'' {\em Progress of Theoretical Physics Supplement}, vol.~172,
  pp.~49--60, 2008.

\bibitem{sperello_2011}
S.~d.~S. {Alighieri}, {\em {Cosmological Birefringence: An Astrophysical Test
  of Fundamental Physics. {\rm Chapter in book "From Varying Couplings to
  Fundamental Physics", Eds. C. Martins and P. Molaro, pp. 139-145} }}.
\newblock Springer-Verlag, Berlin, 2011.

\bibitem{krawex_2009}
M.~{Kramer} and N.~{Wex}, ``The double pulsar system: a unique laboratory for
  gravity,'' {\em Classical and Quantum Gravity}, vol.~26, no.~7, p.~073001,
  2009.

\bibitem{Damour_1991ApJ}
T.~{Damour} and J.~H. {Taylor}, ``{On the orbital period change of the binary
  pulsar PSR 1913 + 16},'' {\em \apj}, vol.~366, pp.~501--511, Jan. 1991.

\end{thebibliography}
\addcontentsline{toc}{section}{\hspace{0.55cm}References}
\end{document}